\documentclass[preprint,12pt]{elsarticle}

\usepackage{graphicx}
\usepackage{amssymb,amsfonts,amsmath}
\usepackage[margin=2.8cm]{geometry}
\usepackage{booktabs}
\usepackage{afterpage}
\usepackage{multirow}
\usepackage{array}

\journal{arXiv}

\begin{document}

\begin{frontmatter}

\title{Apparent criticality and calibration issues\\ in the Hawkes self-excited point process model: \\application to high-frequency financial data}

\author[1]{Vladimir Filimonov}
\ead{vfilimonov@ethz.ch}

\author[1,2]{Didier Sornette}
\ead{dsornette@ethz.ch}

\address[1]{Dept. of Management, Technology and Economics, ETH Z\"{u}rich, Z\"{u}rich, Switzerland}
\address[2]{Swiss Finance Institute, c/o University of Geneva\vspace{-0.8cm}}

\begin{abstract}
{\small   We present a careful analysis of possible issues on the application of the self-excited Hawkes process to high-frequency financial data. We carefully analyze a set of effects leading
to significant biases in the estimation of the ``criticality index'' $n$ that quantifies the
degree of endogeneity of how much past events trigger future events.
We report the following model biases: (i) evidence of strong upward biases on 
the estimation of $n$ when using power law memory kernels
in the presence of outliers, (ii) strong effects on $n$ resulting from the 
form of the regularization part of the power law kernel,
(iii) strong edge effects on the estimated $n$ when using power law kernels, and (iv)
the need for an exhaustive search of the absolute maximum of the log-likelihood function due to its
complicated shape. Moreover, we demonstrate that the calibration of the Hawkes process on
mixtures of pure Poisson process with changes of regime leads to completely spurious apparent
critical values for the branching ratio ($n \simeq 1$) while the true value is actually $n=0$.
More generally, regime shifts on the parameters of the Hawkes model and/or on the
generating process itself are shown to systematically lead to a significant upward bias
in the estimation of the branching ratio. 
We demonstrate the importance of the preparation of the high-frequency 
financial data, in particular: (a) the impact of overnight trading in the analysis 
of long-term trends, (b) intraday seasonality and detrending of the data and 
(c) vulnerability of the analysis to day-to-day non-stationarity and regime shifts. 
Special care is given to the 
decrease of quality of the timestamps of tick data due to latency and grouping of messages
to packets by the stock exchange.
Altogether, our careful exploration of the caveats of the calibration of the Hawkes process 
stresses the need for considering all the above issues before any conclusion can be sustained.
In this respect, because the above effects are plaguing their analyses, the claim by 
Hardiman, Bercot and Bouchaud (2013) that financial market have been continuously
functioning at or close to criticality ($n \simeq 1$) cannot be supported. In contrast, our previous
results on E-mini S\&P 500 Futures Contracts and on major commodity future contracts are upheld.
}
\end{abstract}

\begin{keyword}
Hawkes process, Poisson process, endogeneity, reflexivity, branching ratio, outliers, memory kernel, high-frequency data, 
criticality, statistical biases, power laws, regime shifts
\end{keyword}

\end{frontmatter}

\section{Introduction}
 
The Hawkes self-excited Poisson process is the simplest extension of the Poisson point process,
in which past events influence future events through a memory kernel. 
Its broad domain of applications from biology, geology to economics and finance invites a thorough understanding 
of the issues associated with its calibration to real data and in particular in the possible biases that
arise in the estimation of one of its key parameters, the branching ratio $n$ that quantifies the
degree of endogeneity of how much past events trigger future events. 

We present a careful analysis of a set of effects that lead
to significant biases in the estimation of the branching ratio $n$, arguably the key parameter of the 
Hawkes self-excited Poisson process. The motivation of our study stems from the
meaning of $n$ as a direct measure of endogeneity (or reflexivity), since $n$
is exactly equal to the fraction of  the average number of endogenously generated events among all events 
\citep{Sornette2003geo,FilimonovSornette2012_Reflexivity} for stationary time series.
Concretely, the measure $n=0.7-0.8$ reported in our recent studies
\citep{FilimonovSornette2012_Reflexivity,FilimonovSornetteUNCTAD2012_Commodities}
means that 70 to 80\% of all trades in the  E-mini S\&P 500 Futures Contracts 
and in major commodity future contracts are due in recent years to past trades rather than to
external effects or exogenous news. This result has important implications concerning
the efficient market hypothesis and the stability of financial markets in the presence 
of increasing trading frequency and volume. Our motivation is further increased
by the recent claim based also on the calibration of the Hawkes process that financial market have been continuously
functioning at or close to criticality ($n \simeq 1$) over the last decades \citep{HardimanBouchaud2013},
a result in contradiction with our other studies 
\citep{FilimonovSornette2012_Reflexivity,FilimonovSornetteUNCTAD2012_Commodities}.

The article is structured as follows. In section~\ref{sec:hawkes_endo}, we introduce the Hawkes model and briefly discuss its properties,
and in particular provide the rigorous definition of the branching ratio $n$. We also explain how
the calibration of the Hawkes model to empirical time series is performed and present the residual analysis
as a statistical goodness of fit.  Section \ref{sec:common} discusses the
common issues appearing in the calibration of the Hawkes process, which are divided
in four classes: (i) the impact of outliers, (ii) the somewhat surprising impact of the regularization
part of power law kernels, (iii) the edge effect that is particularly important for long memory
power law kernels and (iv) the often present multiple extrema of the likelihood function. 
Section \ref{sec:microstructure} studies in detail how 
some microstructure patterns of the high-frequency financial data are the source of significant
estimation biases of the branching ratio. In particular, we analyze 
the problem of distinguishing between regular Trading Hours and overnight trading,
the impact of recording latency, of the grouping of timestamps and the
bundling of timestamps. We show that the intraday seasonality leads to a 
non-stationary behavior of the exogenous component of the Hawkes process, which 
is very difficult to remove and is the source of large biases in the estimation of $n$.
The section ends by emphasizing how non-stationarity, regime shifts and the mixing
of different phases leads to extraordinary large spurious calibration results, such as 
a mixture of Poisson processes with $n=0$ by definition for which the calibration
concludes that $n$ is close to critical!  And section 5 concludes by stressing the need to
revisit many previous studies that have been concerned with inter-event times.

\section{Hawkes model and measure of endogeneity}\label{sec:hawkes_endo}

\subsection{Hawkes model and its kernel specification}\label{sec:hawkes}

The methodology for estimating the endogeneity (or reflexivity) present 
in the dynamics of a given point process, developed in \citep{FilimonovSornette2012_Reflexivity} and later exploited in~\citep{FilimonovSornetteUNCTAD2012_Commodities,HardimanBouchaud2013,FilimonovWeatleySornette2012_ACD}, is based on the self-excited conditional Poisson model introduced by  \cite{Hawkes1971_orig,Hawkes1971}.
Being a point process, the linear Hawkes model is defined with the conditional intensity
\begin{equation}\label{eq:lamdba_conditional}
	\lambda(t|\mathcal{F}_{t-})=\lim_{h\downarrow 0}\frac{1}{h}\mathrm{Pr}\big[N(t+h)-N(t)>0|\mathcal{F}_{t-}\big],
\end{equation}
where  $\{t_{i}\}_{i\in\mathbb{N}}$ is an ordered set of event times ($t_i\leq t_j$ for $i<j$); $N(t)=\max(i:t_{i}\leq t)$ is the corresponding \emph{counting process} and $\mathcal{F}_{t-}=\{t_{1},\dots,t_{i}:t_i<t\}$ is the \emph{filtration} that represents the history of the process until time $t$.
Defining $\mu(t)$ as the \emph{background intensity}, which is a deterministic function of time that accounts for the intensity of arrival of \emph{exogenous} events (not dependent on history), the conditional intensity
of the Hawkes process takes the following general form:
\begin{equation}\label{eq:hawkes_general}
	\lambda(t|\mathcal{F}_{t-})=\mu(t)+\int_{-\infty}^{t}\varphi(t-s)dN(s),
\end{equation}
A deterministic  \emph{kernel function} $\varphi(t)$, which should satisfy causality ($\varphi(t)=0$ for $t<0$), models the \emph{endogenous} feedback mechanism (memory of the process). The integral of $\varphi(t)$, which is called the 
branching ratio,
\begin{equation}\label{eq:n}
	n := \int_0^\infty \varphi(t)dt>0,
\end{equation}
plays a crucial role for the dynamics of the process, which will be elaborated later. In particular,
stationarity of the Hawkes process~\eqref{eq:hawkes_general} requires that $n \leq 1$. To emphasize the importance of this parameter, we rewrite~\eqref{eq:hawkes_general} as
\begin{equation}\label{eq:hawkes_discr}
	\lambda(t|\mathcal{F}_{t-})=\mu(t)+n\sum_{t_i<t} h(t-t_i),
\end{equation}
where we have also accounted for the fact that each event arrives instantaneously and the differential of the counting process $dN(t)$ can be represented in the form of a sum of delta-functions $dN(t)=\sum_{t_i<t}\delta(t-t_i)dt$. 
Here, $h(t)$ is the normalized kernel function $h(t) =\varphi(t)/n$, such that $\int_0^\infty h(t)dt=1$.

The shape of the kernel function $h(t)$ defines the correlation properties of the process. Financial applications
traditionally use an exponential kernel 
\citep{Hewlett2006,Bowsher2007,Cont2011,FilimonovSornette2012_Reflexivity,FilimonovSornetteUNCTAD2012_Commodities} 
\begin{equation}\label{eq:exp}
	h(t)=\frac1\tau \exp\left(-\frac{t}{\tau}\right)~\chi(t)~.
\end{equation}
This exponential form has been originally suggested by \cite{Hawkes1971_orig} and ensures Markovian properties of the model~\citep{Oakes1975}. The Heaviside function $\chi(t)$ ensures the 
validity of the causality principle. In the geophysical applications of the Hawkes model, in the form
of its spatio-temporal extension called the \emph{Epidemic-Type Aftershock sequence (ETAS)}~\citep{VereJonesOzaki1982,VereJones1970,Ogata1988,Sornette2002geo_regimes}, the memory kernel $h(t)$
has a power law time-dependence:
\begin{equation}\label{eq:pow}
	h(t)=\frac{\theta c^\theta}{(t+c)^{1+\theta}}~\chi(t),
\end{equation}
which describes the modified Omori-Utsu law of aftershock rates~\citep{Utsu1961,UtsuOgata1995}. 
The time constant $c$ regularizes the behavior of the power law kernel at very short times.
Kagan and Knopoff introduced another regularization for the memory kernel
\citep{Kagan1981,Kagan1987}
\begin{equation}\label{eq:pow_jpb}
	h(t)=\frac{\epsilon \tau_{0}^\epsilon}{t^{1+\epsilon}}~\chi(t-\tau_0)~.
\end{equation}
This expression (\ref{eq:pow_jpb}) was also used recently in \citep{HardimanBouchaud2013},
and was approximated as a sum of exponential functions,
\begin{equation}\label{eq:pow_jpb_sum}
	h(t)=\frac1Z\left[
	\sum_{i=0}^{M-1}\frac1{\xi_i^{1+\epsilon}}\exp\left(-\frac{t}{\xi_i}\right) - 
	S\exp\left(-\frac{t}{\xi_{-1}}\right),
	\right]
\end{equation}
where the coefficients obey a power law  $\xi_i=\tau_0 m^i$ and the coefficients $S$ and $Z$ are chosen so that $h(0)=0$ and $\int_0^\infty h(t)dt=1$. In their empirical calibration, \cite{HardimanBouchaud2013} has fixed $M=15$ and $m=5$. For values of $\epsilon$ close to zero, the resulting function describes approximately a power-law form with tail exponent $1+\epsilon$, while the negative exponential term provides a smooth cut-off at short times.

\subsection{The branching ratio}\label{sec:branching}

The linear structure of the intensity of the Hawkes process~\eqref{eq:hawkes_discr} allows one to consider it as a cluster process in which the random process of cluster centers $\{t^{(c)}_i\}_{i\in\mathbb{N}_{>0}}$ is the Poisson process with rate $\mu(t)$. All clusters associated with centers $\{t^{(c)}_i\}$ are mutually independent by construction and can be considered as a \emph{generalized branching process}~\citep{Hawkes1974}. In this context, each event $\{t_i\}$ can be either an \emph{immigrant} or a \emph{descendant}. The rate of immigration is determined by the background intensity $\mu(t)$ and results in an exogenous random process. Once an immigrant event occurs, it generates a whole cluster of events. Namely, a zeroth-order event (which we will call the \emph{mother event}) can trigger one or more first-order events (\emph{daughter events}). Each of these daughters, in turn, may trigger several second-order events (the grand-daughters of the initial mother), and so on. 

In this context, the \emph{branching ratio} $n$ is defined as the average number of daughter events (i.e. 
triggered events of first generation) per mother event. Depending on the branching ratio, there are three regimes: (i) \emph{sub-critical} ($n<1$), (ii) \emph{critical} ($n=1$) and (iii) \emph{super-critical} or explosive ($n>1$). Starting from a single mother event (or immigrant) at time $t_1$, the process dies out with probability $1$ in the sub-critical and critical regimes and has a finite probability to explode to an infinite number of events in the super-critical regime. The critical regime for $n=1$ separates the two main regimes and is characterized by power law statistics of the number of events and in the number of generations before extinction~\citep{SaichevSornette2005}. For $n\leq 1$, the process is stationary in the presence of a Poissonian or more generally stationary flux of immigrants.

In the case of a constant background intensity ($\mu(t)=\mu=\mbox{const}$) and in 
the sub-critical regime ($n<1$), the branching ratio is exactly equal to the average fraction 
of the number of descendants in the whole population of events
\citep{Sornette2003geo,FilimonovSornette2012_Reflexivity}. In other words, the branching ratio
is equal to the average proportion of endogenously generated events among all events and can be considered as an effective measure of endogeneity of the system.

\subsection{Estimation of the degree of endogeneity $n$} 

There are several routes to estimate $n$ from real data. One is to reverse-engineer the clusters and calculate the average number of direct descendants to any given event. This can be done via the stochastic declustering (parametric~\citep{Zhuang2002} and non-parametric~\citep{Marsan2008,LewisMohler2011}) method, which amounts to reconstruct from the sequence of events the original cluster (branching) structure, or at least distinguish between descendants and immigrants, but this may have severe limitations in the presence of long-memory kernels~\citep{SornetteUtkin2009}. A simpler way is to just use the Maximum Likelihood Estimation method, which benefits from the fact that the log-likelihood function is known in closed form for Hawkes processes \citep{Ogata1978,Ozaki1979}. Namely the parameters of the model~\eqref{eq:hawkes_discr} with any specified kernel~\eqref{eq:exp}--\eqref{eq:pow_jpb_sum} can be determined by numerical maximization of the following log-likelihood function:
\begin{equation}\label{eq:loglik}
	\log L(t_1,\dots,t_N)=-\int_0^T \lambda(t|\mathcal{F}_{t-})dt+
	\sum_{i=1}^N\log \lambda(t_i|\mathcal{F}_{t_i-}),
\end{equation}
where $\{t_i\in(0,T]\}$ is the set of observation times of the events.  In general, the calculation of the log-likelihood function~\eqref{eq:loglik} has computational complexity $\mathcal{O}(N^2)$. However, for exponential~\eqref{eq:exp} and approximate power law~\eqref{eq:pow_jpb_sum} kernels, it can be reduced to $\mathcal{O}(N)$ by taking advantage of a recursive relation~\citep{Ozaki1979}.

It should be noted that the calibration of the Hawkes model on finite samples always results in 
an underestimation of the real value of the branching ratio $n$. Indeed, 
the events occurring before the time window of calibration that could trigger
events within the window are not taken into account. This results in an overestimation of the background rate $\mu$ and therefore an underestimation of the observed $n$. In other words, neglecting past events before the windows and their triggering effect leads to the misattribution that many of the endogenous events are exogenous~\citep{Sornette2005geo_clustering}. Moreover, for any given event, not all its daughter events are observed within the given window $(0, T]$, especially for mother events that happen to be close to the right-end boundary $T$. This effect, which also results in underestimating the observed secondary events and thus $n$, becomes 
more pronounced for larger memories of the process, as determined by the shape
and characteristics time of the kernel function $h(t)$.

Finally, the calibration of the Hawkes model on the data should be validated with the goodness-of-fit using
 \emph{residual analysis}~\citep{Ogata1988}, which consists in studying the \emph{residual process}, defined as the nonparametric transformation of the initial series of the event time stamps $t_{i}$ into 
\begin{equation}
	\xi _{i}=\int_{0}^{t_{i}}\hat{\lambda}_{t}(t)dt,  \label{eq:xi}
\end{equation}%
where $\hat{\lambda}_{t}(t)$ is the conditional intensity of the Hawkes
process~\eqref{eq:hawkes_discr} estimated with the maximum likelihood method. As it
was shown in~\citep{Papangelou1972}, under the null hypothesis that the data
has been generated by the Hawkes process~\eqref{eq:hawkes_discr} with selected kernel~$h(t)$, the residual process $\xi _{i}$ should be Poisson with unit intensity. The goodness-of-fit can then be verified both by (i)
visual cusum plot or Q-Q plot analysis and (ii) rigorous statistical tests, such as independence tests applied to the sequence of $\xi _{i}$ and/or tests of the exponential distribution of the transformed inter-event times $\xi _{i}-\xi _{i-1}$, which amounts to testing the uniform distribution of the random variables $U_{i}=1-\exp [-(\xi _{i}-\xi _{i-1})]$ in the interval $[0,1]$). The null hypothesis of uniform probability distribution of variables $U_i$ can be tested using the Kolmogorov-Smirnov test and the null hypothesis of absence of autocorrelations in sequence $U_1,\dots,U_{N-1}$ can be addressed with the Ljung-Box test.

\subsection{Implementation of the Maximum Likelihood Estimation method \label{numerical}}

Consider the Maximum Likelihood Estimation of the Hawkes process~\eqref{eq:hawkes_discr}, where the background intensity is constant ($\mu(t)=\mu=\mbox{const}$) and the kernel $h_{\psi}(t)$ is parametrized with a parameter set~$\psi$ ($\psi=\{\tau\}$ for the exponential kernel~\eqref{eq:exp}, $\psi=\{c,\theta\}$ for the Omori-type kernel~\eqref{eq:pow} and $\psi=\{\tau_0,\epsilon\}$ for the exponential kernel~\eqref{eq:pow_jpb_sum}). The expression for the log-likelihood function~\eqref{eq:loglik} in this case can be written in the form:
\begin{equation}\label{eq:loglik_discr}
	\log L(t_1,\dots,t_N)=-\mu T-n H_1(\psi)+\sum_{i=1}^N\log \big(\mu+n H_2(\psi, t_i)\big),
\end{equation}
where
\begin{equation}\label{eq:H12}
	H_1(\psi) = \sum_{i=1}^N\int_0^T h_{\psi}(t-t_i)dt,\qquad
	H_2(\psi, t_i) = \sum_{t_j<t_i}h_{\psi}(t_i-t_j).
\end{equation}
The most computationally intensive part of the log-likelihood calculation is the summation over all past events in $H_2(\psi,t_i)$. We propose here to partition the search space into two subspaces and to subordinate one to the other, as demonstrated recently in another similar application~\citep{FilimonovSornette2011_LPPL_calibration}. 
We thus reformulate the optimization problem
\begin{equation}\label{eq:optim_0}
	\{\hat\mu,\hat n,\hat\psi\} = \arg\min_{\mu,n,\psi}\Big[-\log L(\mu,n,\psi|t_1,\dots, t_N)\Big]
\end{equation}
into the two-step optimization problems:
\begin{equation}\label{eq:optim_1}
	\hat\psi = \arg\min_{\psi} S(\psi|t_1,\dots, t_N),
\end{equation}
where
\begin{equation}\label{eq:optim_2}
\begin{array}{rcl}
	S(\psi|t_1,\dots, t_N) &=& \displaystyle\min_{\mu,n} \Big[-\log L(\mu,n,\psi|t_1,\dots, t_N)\Big]
\\
&=& \displaystyle\min_{\mu,n} \left[\mu T+n H_1(\psi)-\sum_{i=1}^N\log \big(\mu+n H_2(\psi, t_i)\big)\right].
\end{array}
\end{equation}
In other words, the cost function $S(\psi|t_1,\dots, t_N)$ 
of the parameters set $\psi$ of kernel function $h_\psi(t)$ is equal to the value of 
the original cost function ($-\log L$) when the parameters $\{\mu,n\}$ are selected as the best ones for a given value of $\psi$.
This reformulation achieves three important goals: (i) similarly to~\citep{FilimonovSornette2011_LPPL_calibration}, in some cases, it decreases the number of local minima; 
(ii) it allows us to present an illustrative visualization of the search space (see Section~\ref{sec:extrema}) and 
(iii) it dramatically decreases the computational cost of the calibration by using dynamic programming.
Fixing the parameter set $\psi$ of the kernel function allows us to 
solve efficiently the optimization problem~\eqref{eq:optim_2} by
computing $H_1(\psi)$ and $H_2(\psi,t_i)$ only once for each step of problem~\eqref{eq:optim_1}.  

Further, in the subordinated optimization over the space $\{\mu,n\}$ we can simplify~\eqref{eq:optim_2} by 
an analytical determination of one of the parameters using the method proposed by Lyubushin and Pisarenko~\cite{LyubushinPisarenko1994}. This method is based on the fact that the log-likelihood~\eqref{eq:loglik_discr} satisfies the following equation:
\begin{equation}\label{eq:part_diff}
   \mu\frac{\partial \log L}{\partial \mu}+  n\frac{\partial \log L}{\partial n}=-\mu T-n H_1(\psi)+N,
\end{equation}
where $N$ is the total number of observed events within $(0,T]$. At the extremum of the log-likelihood (for $\mu=\hat\mu$, $n=\hat n$ and $\psi=\hat\psi$), the partial derivatives are vanishing ($\partial \log L/\partial \mu=\partial \log L/\partial n=0$) and equation~\eqref{eq:part_diff} provides a relation between the optimal parameter values:
\begin{equation}\label{eq:mu_n_psi}
	\hat\mu T+\hat n H_1(\hat \psi)=N.
\end{equation}
Taking into account this last expression (\ref{eq:mu_n_psi}), one can replace the optimization problem~\eqref{eq:optim_2} by the equivalent problem:
\begin{equation}\label{eq:optim_3}
	S(\psi|t_1,\dots, t_N) = 
 \min_{n} \left[N-\sum_{i=1}^N\log \left(\frac{N}{T}+n \frac{T H_2(\psi, t_i)-H_1(\psi)}{T}\right)\right].
\end{equation}
Despite the fact that the profiles of the cost functions in~\eqref{eq:optim_3} and~\eqref{eq:optim_2} are different, they have the same minimal value. The optimization procedure now consists of the subordinated optimization~\eqref{eq:optim_1} and~\eqref{eq:optim_3} that yields the values of $\hat\mu$ and $\hat\psi$,  while the value of $\hat\mu$ is derived from~\eqref{eq:mu_n_psi}: \mbox{$\hat\mu=N/T-\hat nH_1(\hat\psi)/T$}.

Note that the quality of the parameter estimation depends on the specific parametrization of the model~\eqref{eq:hawkes_general}. The classical econometric literature, such as~\citep{Bowsher2007,Bauwens2009,Errais2010,AitSahalia2011,Embrechts2011}, employ the Hawkes model with the exponential kernel $\varphi(t)=\alpha\exp(-\beta t)$, which is not normalized and require the estimation of the two parameters $\{\alpha, \beta\}$, which just have to obey the conditions \mbox{$\alpha, \beta>0$}. In contrast, the explicit definition of the branching ratio $n=\alpha/\beta$ (see Eq.~\eqref{eq:hawkes_discr}), which is bounded  for stationary process $(0\leq n\leq 1)$, provides much more robust estimations. For similar reasons, in the subordinated procedure~\eqref{eq:optim_2}, we suggest inferring the background activity (\mbox{$\mu>0$}) from~\eqref{eq:mu_n_psi}, while minimizing the modified cost function with respect to $n$ as in~\eqref{eq:optim_3}, and not vice versa.

\clearpage
\section{Common issues of the calibration of the Hawkes process \label{sec:common}}

In this and following sections, we discuss a number of issues arising in the treatment of high frequency financial data and in numerical procedures for the calibration of the Hawkes model that bias the estimation of the parameters of the model. Some of these issues are common to all types of problems, and some are specific to the work~\citep{HardimanBouchaud2013},
which claims that financial markets are functioning systematically at criticality. We document
and quantify precisely a series of biases associated with the estimation of the branching
ratio $n$ that rationalize the spurious claim: many effects concur to give the impression of an apparent
criticality and these effects need to be understand, identified and corrected before any solid
conclusion can be drawn.

\subsection{Impact of outliers for different memory kernels}\label{sec:robustness}

The first issue concerns the robustness of the estimation of the branching ratio $n$ in the presence
of a small fraction of outliers and for different memory kernels. This problem is motivated by 
the observed distribution of inter-quote durations for mid-quote price changes of E-mini 
S\&P 500 Futures Contracts during Regular Trading Hours (from 9:30 to 16:15 CDT), which forms the basis of the 
contradictory claims presented in \citep{HardimanBouchaud2013} 
versus \citep{FilimonovSornette2012_Reflexivity}.

By comparing Tables~\ref{tb:intervals_empirical} and~\ref{tb:intervals_model}, we observe the existence of 
genuine outliers in the data, as compared
with what is expected from time series generated purely with the Hawkes process. 
Specifically, we can observe that the relation between quantiles of inter-quote durations (e.g. fractions $Q_{95}/Q_{90}$ or $Q_{99}/Q_{95}$) is similar for theory (Table~\ref{tb:intervals_model}) and empirical data (Table~\ref{tb:intervals_empirical}). However the relation of the maximal observed duration to the quantiles ($Max/Q_{99}$) is different. When, in theory, the maximum observed inter-quote duration is only $\sim 3$ times larger than $Q_{99}$, the empirical observations show that the maximum observed inter-quote duration is more than ten times larger than the 99\% quantile ($Q_{99}$) even in the years of active trading.
This reflects a highly irregular trading activity in the market that is not fully captured at the extremes
by the best Hawkes process calibrated on the same data.

\begin{table}[t!]
\caption{Descriptive statistics of inter-quote durations (between consecutive mid-quote price changes during Regular trading Hours) of E-mini S\&P 500 Futures Contracts in different time periods: empirical quantiles ($Q_{90}$, $Q_{95}$, $Q_{99}$), maximum value (Max), total number of observed mid-quote durations (N), number of durations that are at least twice greater than $Q_{99}$ (N$_{>2Q_{99}}$) and fraction of latest in the overall sample (Fraction$=\mbox{N}_{>2Q_{99}}/\mbox{N}\cdot100\%$). }
\begin{center}
\begin{tabular}{ccccccccc}
\toprule
Date from & Date to & $Q_{90}$ & $Q_{95}$ & $Q_{99}$ & Max & N & N$_{>2Q_{99}}$ & Fraction \\ 
\toprule
February 1, 2002 & April 1, 2002  & 13.4 & 20.1 & 40.5 & 458.9 & 161573 & 209 & 0.13\% \\
February 1, 2006 & April 1, 2006 & 23.3 & 39.2 & 88.5 & 933.1 & 119193 & 172 & 0.14\% \\
February 1, 2009 & April 1, 2009 & 5.1 & 8.5 & 19.1 & 330.0 & 454499 & 795 & 0.17\% \\
February 1, 2012 & April 1, 2012 & 4.2 & 10.8 & 38.6 & 888.0 & 313436 & 1191 & 0.38\% \\
\bottomrule
\end{tabular}
\end{center}
\label{tb:intervals_empirical}
\caption{Theoretical quantiles and maximum values of inter-event durations for time series generated with the Hawkes process with the approximate power law kernel~\eqref{eq:pow_jpb_sum} for $\mu=0.02$, $\epsilon=0.15$, and $n$ and $\tau_0$ given in the first two columns (parameters are taken from \citep{HardimanBouchaud2013}). The data is obtained by numerical simulation of the Hawkes process on the interval $(0, 10^8+10^5]$ with burning of the interval $(0, 10^8].$}
\begin{center}
\begin{tabular}{ccccccc}
\toprule
n & $\tau_0$ & & $Q_{90}$ & $Q_{95}$ & $Q_{99}$ & Max \\ 
\toprule
\multirow{3}{*}{0.3} & 1 & & 92.7 & 122.6 & 199.1 & 372.0 \\
& 0.1 & & 88.7 & 115.8 & 195.0 & 390.6 \\
& 0.01 & & 92.9 & 124.8 & 203.5 & 381.5 \\
\hline
\multirow{3}{*}{0.5} & 1 & & 66.2 & 90.8 & 151.2 & 261.8 \\
& 0.1 & & 66.0 & 91.5 & 157.2 & 284.6 \\
& 0.01 & & 71.7 & 98.9 & 163.1 & 299.2 \\
\hline
\multirow{3}{*}{0.7} & 1 & & 41.0 & 58.7 & 105.1 &  226.7 \\
& 0.1 & & 45.7 &  67.2 & 116.2 & 247.7 \\
& 0.01 & & 45.3 & 66.6 & 120.8 & 254.5 \\
\hline
\multirow{3}{*}{0.95} & 1 & & 10.3 & 14.9 & 26.0 & 65.2 \\
& 0.1 & & 9.8 & 14.8 & 28.2 & 80.4 \\
& 0.01 & & 8.5 & 14.2 & 29.2 & 89.0 \\
\hline
\multirow{3}{*}{0.99} & 1 & & 4.9 &  7.1 & 12.4 & 33.9 \\
& 0.1 & & 3.3 &  4.8 &  9.2 & 31.3 \\
& 0.01 & & 1.9 &  3.2 &  6.6 &  23.8\\
\bottomrule
\end{tabular}
\end{center}
\label{tb:intervals_model}
\end{table}

To illustrate potential biases that result from this effect,
we introduce a few outliers (extreme inter-event intervals) in synthetic time series
generated by the Hawkes process, so as to mimic the phenomenon observed in Table~\ref{tb:intervals_empirical}
compared with Table 2. We create different synthetic time series of the Hawkes process,
with duration $(0, 10^5+10^4]$ seconds and fixed exogenous intensity $\mu=0.3$ and branching ratio $n=0.7$ using
\begin{itemize}
\item[(i)] the exponential kernel~\eqref{eq:exp} with $\tau=0.1$ or
\item[(ii)] the power law kernel~\eqref{eq:pow} with $c=0.1$ and $\theta=0.5$ or
\item[(iii)] the approximate power law kernel~\eqref{eq:pow_jpb_sum} with $\tau_0=0.1$ and $\epsilon=0.5$.
\end{itemize}
In order to get rid of the edge effects, we burn the initial period $(0, 10^5]$ seconds 
(we discuss the impact of the edge effect in details in section~\ref{sec:edge}). 
In each synthetic time series, we record the maximum observed duration, and then replace a small fraction of randomly selected durations with values that are M-times ($M=1,2$ and 5) larger than the initial maximum observed value. On these time series with a small fraction 
of outliers, we calibrate the Hawkes model with the same kernel as the one used to initially
generate the synthetic time series. This is repeated 100 times to obtain a statistical average and standard
deviation of the estimated branching ratio $\hat n$. 

\begin{figure}[t!]
  \centering
  \includegraphics[width=0.8\textwidth]{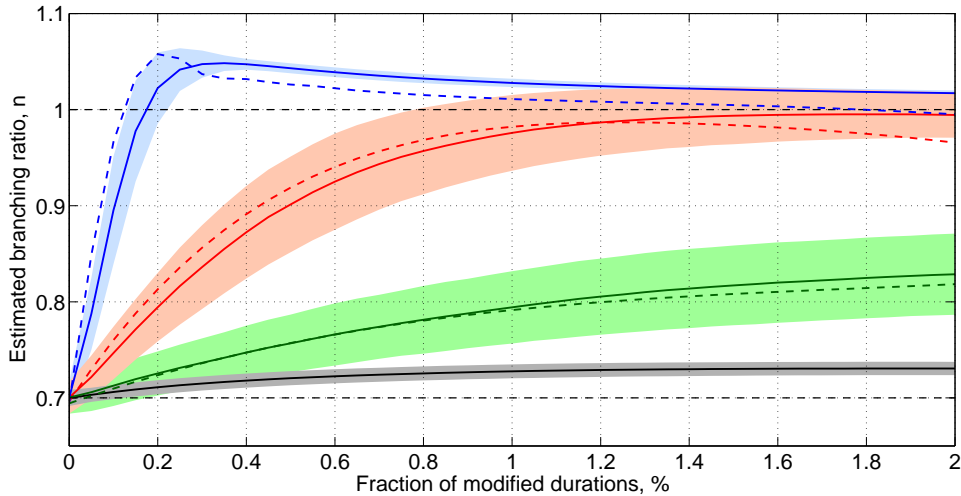}
  \caption{Estimated branching ratio $n$ obtained by calibrating the Hawkes model on synthetic
 time series  generated with the Hawkes process for the three different memory kernels discussed
 in the text, in the presence of a small fraction of inter-quote duration outliers.
 The outliers are generated by replacing a fraction (given in the abscissa) of the durations 
 by durations that are $M$ times larger than the largest duration of the simulation, where $M=1$ (green lines), $M=2$ (red lines) and $M=5$ (blue lines). The lower black continuous line corresponds to the exponential kernel \eqref{eq:pow} with $M=5$.
 Solid blue and red lines correspond to the approximate power law kernel~\eqref{eq:pow_jpb_sum} and dashed blue and red lines correspond to the power law kernel \eqref{eq:pow}.
 Shaded areas cover the $\pm$ one standard deviation of the statistical estimation of $\hat n$ over 100 realizations.
The horizontal dashed line at $n=0.7$ is the true value used in the synthetic generation
of all time series. The critical value  $n=1$ is also indicated as a horizontal dashed line.} 
\label{fig:robustness}
\end{figure}

The results are shown in Figure~\ref{fig:robustness}, which gives the estimated branching ratio
as a function of the fraction of introduced outliers for the three types of memory kernels.
One can observe that the estimations of the time series generated with an exponential kernel are robust to the introduction
of outliers, as the estimated $\hat n$ remains within one standard deviation of the true value $0.7$ 
used to generate the synthetic time series. Remarkably, a completely different behavior
occurs for power law kernels, for which the estimated branching ratio is strongly biased upward even for small fraction of outliers. So that for $M=2$, $1\%$ of outliers introduce an upward bias of approximately $0.28$ (or 40\%
in relative value), resulting in an almost critical estimated $\hat n \approx 0.98$.
This bias is much stronger for $M=5$, where just $0.17\%$ of outliers are sufficient
to lead to the spurious conclusion that the system is at or close to critical ($\hat n \simeq 1$) or
even super-critical, in the case of exact power law kernels. 
The intuition behind this result is that large outlier time intervals are ``interpreted'' 
incorrectly within the Hawkes model with power law kernel as waiting times that reflect
a genuine endogenous triggering activity. The scale-free power law kernels are flexible enough
to ``endogenize'' these outliers. In contrast, the more rigid form of the exponential kernel, 
which is characterized by a single characteristic time scale $\tau$, leads to a much smaller
influence of the outliers in the calibration.

In real data, as we observe from the Table~\ref{tb:intervals_empirical}, approximately 0.1--0.4\% of inter-event durations are at least twice larger than the 99\% quantile. Though the exact value of the bias for $\hat n$ depends on the distribution and location of extreme durations, the synthetic case for $M=1$ can be considered as a conservative estimate of the bias with $M=2$ providing a reasonable order-of-magnitude value.

\subsection{Effect of the regularization part of power law kernels}

\begin{figure}[h!]
  \centering
  \includegraphics[width=0.8\textwidth]{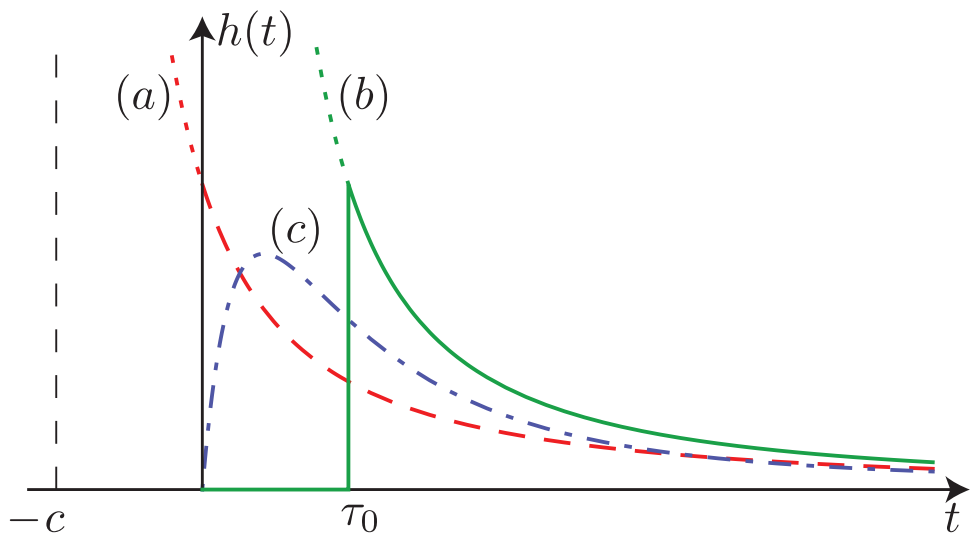}
  \caption{Illustration of the differences between the three power law kernels 
  implemented within the Hawkes model: (a) the Omori law~\eqref{eq:pow} (continuous green line),
  (b) the power law kernel with a cut-off~\eqref{eq:pow_jpb} (dashed red line) and (c) the approximation 
  of the power law kernel with a sum of exponents~\eqref{eq:pow_jpb_sum} (dotted-dashed blue line).} \label{fig:kernels}
\end{figure}

In section~\ref{sec:hawkes} presenting the Hawkes model~\eqref{eq:hawkes_discr},
we have introduced three different versions of the memory kernel $h(t)$ with power law tail,
which differ in the way the regularization at short times is introduced as shown in figure \ref{fig:kernels}:
\begin{enumerate}[(a)]
\item the Omori law~\eqref{eq:pow} with a smooth regulation, 
\item the power law kernel with a sharp ultraviolet cut-off~\eqref{eq:pow_jpb} and 
\item the approximation of the power law kernel with a sum of exponentials~\eqref{eq:pow_jpb_sum}.
\end{enumerate}
While their tail are identical asymptotically, they differ significantly at short times. 
In particular, while the Omori law kernel is strictly decaying from $t=0$, the other two 
are non-monotonous and are characterized by a maximum at some $t_{max}>0$, corresponding
to the most probable waiting time between a mother and its first-generation triggered daughters.

These apparently innocuous differences have actually a significant impact on the estimation
of the branching ratio $n$, leading to important biases when the kernel is not specified correctly. In order to illustrate this, we have numerically simulated the Hawkes process with one of the power law kernels presented in Fig.~\ref{fig:kernels} and calibrated the Hawkes model with another of these power law kernel on this synthetic data. We have considered the following cases:
\begin{enumerate}[(i)]
\item we simulated the Hawkes process with the approximate power law kernel~\eqref{eq:pow_jpb_sum} for $\tau_0=1$ and $\epsilon=0.5$ and calibrated the obtained time series with the Hawkes process with the Omori law kernel~\eqref{eq:pow};
\item we simulated the Hawkes process with the Omori law kernel~\eqref{eq:pow} for $c=1$ and $\theta=0.5$ and calibrated the obtained time series with the Hawkes process with the approximate power law kernel~\eqref{eq:pow_jpb_sum};
\item finally, to assess the possible bias of the estimation procedure, we simulated the Hawkes process with the approximate power law kernel~\eqref{eq:pow_jpb_sum} for $\tau_0=1$ and $\epsilon=0.5$ and calibrated 
the obtained time series with the Hawkes process with the same kernel.
\end{enumerate}
We have fixed the background activity $\mu=0.1$ and spanned the branching ratio $n$ in the interval $[0, 1]$. In order to get rid of edge effects, we simulated time series with duration $(0, 10^8+10^5]$ seconds and burned the initial period $(0, 10^8]$ seconds to only analyze the interval $(10^8+1, 10^8+10^5]$.

\begin{figure}[h!]
  \centering
  \includegraphics[width=0.6\textwidth]{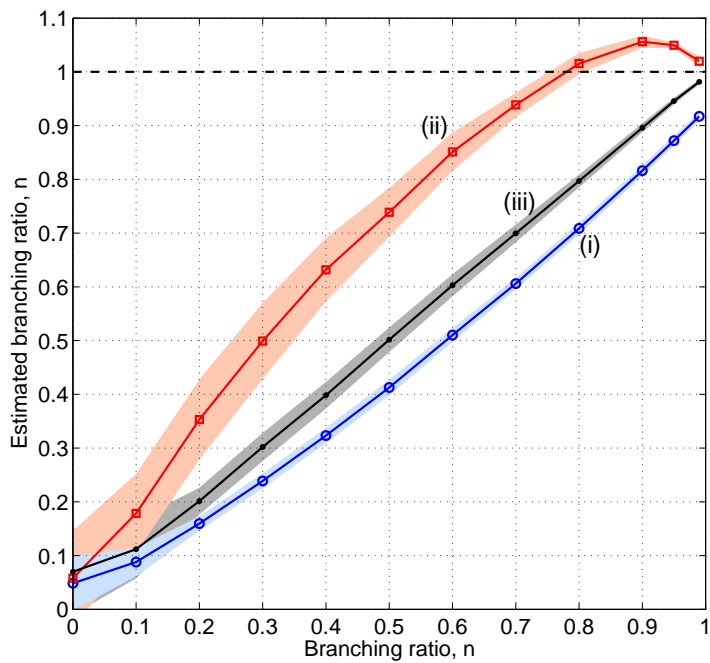}
  \caption{Illustration of the bias in estimating the branching ratio, which is caused by the misspecification of the long-memory kernel (see text). (i) Estimated branching ratio $\hat n$ using the Hawkes model with the Omori law kernel~\eqref{eq:pow}, when the generating process has the approximate power law kernel~\eqref{eq:pow_jpb_sum} (blue curve with circles); (ii) estimated branching ratio using the Hawkes model with the approximate power law kernel, when the generating process has the Omori law kernel (red curve with squares); (iii) estimated branching ratio in the case when both generating model and estimating model have the approximate power law kernel (black curve with dots). Solid lines show the average value of $\hat n$ obtained over 100 estimations. The shaded areas cover the $\pm$ one standard deviation of the statistical estimation over the same 100 realizations.} \label{fig:pow_to_jpb}
\end{figure}

Figure \ref{fig:pow_to_jpb} shows the results of the numerical estimations of $\hat n$. The straight diagonal line and the narrow confidence bands for case (iii) confirms the excellent quality of the calibration in the case of a correct specification of the Hawkes model. However when the memory kernel of the generating process differs from the kernel used in the calibration procedure, significant biases are observed. In case (i) when the true kernel is the approximate power law kernel~\eqref{eq:pow_jpb_sum} and the calibration procedure uses the Omori law~\eqref{eq:pow},
 we observe a slight ($n-\hat n\lesssim 0.07$) underestimation of the branching ratio. However,
 in the opposite case (ii) when the true kernel is the Omori law and the estimation is performed using the approximate power law kernel~\eqref{eq:pow_jpb_sum}, the overestimation of the branching ratio is large, of
 the order of $\hat n-n\gtrsim 0.2$ for  $n>0.3$. In fact, the misspecification of the kernel in this case may lead to 
 the incorrect conclusion of criticality, when the real branching ratio is subcritical ($n\approx0.8$).

The above observations suggest that the choice of the kernel should be a subject of a careful analysis
for any empirical calibration in which a long-memory kernel is used.
Our above tests also challenges the claims of \citep{HardimanBouchaud2013}, since
they have been based on the use of  the approximate power law kernel~\eqref{eq:pow_jpb_sum}
for the calibration of the empirical data, which leads to the largest upward bias for $n$,
even leading to spurious criticality.

The question of the proper form of the power law kernel is not that trivial. For example, even in the synthetic cases
discussed here, we find that residual analysis is usually unable to reject the false model in the sub-critical regime, especially far from criticality ($n<0.7$). Nested statistical tests~\citep{Wilks1938} are not applicable here, since none of the models \eqref{eq:pow}--\eqref{eq:pow_jpb_sum} can be embedded into another one. We find that the Akaike information criterion~\citep{Akaike1974} can successfully select the correct model in our synthetic cases. However, the usefulness
and selectivity of the Akaike criterion for the Hawkes model in the presence of noise remains open
and should be further investigated. 

More specific ways of recovering the kernel involve non-parametric estimation methods, which however
also exhibit severe limitations.  The parametric and non-parametric Expectation Maximization (EM) methods~\citep{Zhuang2004,Marsan2008,VeenSchoenberg2008,LewisMohler2011} typically penalize irregularity of estimated functions to avoid large fluctuations, which will pose a problem when the kernel of the Hawkes process is a power law with a cut-off~\eqref{eq:pow_jpb} or the approximated power law~\eqref{eq:pow_jpb_sum} with a small value of $\tau_0$.
Moreover, the power of these methods is small when the clusters of the Hawkes process overlap significantly~\citep{SornetteUtkin2009}, which is typically the case with financial high frequency data. 

The recently proposed nonparametric method \citep{BacryMuzy2012}, based on the estimation of 
the autocorrelation function of the counting process $dN(t)$, which is also used in~\citep{HardimanBouchaud2013}, is free of this limitation. However, its numerical implementation faces severe short-comings when implemented in short time windows. The counting process is bursty by nature and its autocorrelation function is long-ranged, mimicking the long-memory property of financial volatility. Therefore, its estimation on short intervals is strongly biased. Moreover, since it does not decay to zero within the interval of observation, the Discrete Fourier Transform of the sample will be contaminated by high frequencies because of the truncation (the so-called spectrum leakage). Because 
the Fourier spectrum of the correlation function is transformed into a nonlinear
formula for the Fourier spectrum of the kernel, higher harmonics of 
the spectrum of the kernel, which are responsible for the behavior of the kernel $h(t)$ for small values of $t$,
appear due to a nonlinear transformation of the higher harmonics of the spectrum of the correlation function (see for instance Eq.~(33) in \citep{BacryMuzy2012}).
These higher harmonics are in general not estimated reliably in comparison with the lower frequencies,
which themselves are responsible for the long-term behavior of the kernel $h(t)$. Enlarging
the time window in this case is not generally possible due to the presence of strong intraday non-stationarity. The problem of estimating the kernel at short lags is illustrated in Fig.~1 of \citep{BacryMuzy2012}, where the strictly decaying exponential kernel~\eqref{eq:exp} was estimated with the condition of having a narrow ``cut-off'' at short lags.

Finally, in financial applications, all methods described above are seriously affected by a strong noise due to the nature of the financial data feeds. As we will discuss in section~\ref{sec:millisecond}, even though the timestamps have millisecond resolution, the effective resolution of the data is much lower (up to seconds in early 2000s), which makes all estimations at short lags particularly unreliable.

\subsection{The edge effect and issues concerning numerical simulations of long memory processes}\label{sec:edge}

For the exponential kernel~\eqref{eq:exp} and power law kernels 
\eqref{eq:pow}--\eqref{eq:pow_jpb_sum} with sufficiently large exponents $\theta\ge 1$ 
and $\epsilon\ge 1$, the decay of the kernel as time 
increases is sufficiently fast so that transient periods 
affecting simulations are relatively short. However in the case of small values of the 
exponents $\epsilon$ and $\theta$, the decay of the power law kernels~\eqref{eq:pow} 
and~\eqref{eq:pow_jpb_sum} is very slow, which implies
that events far in the past continue to influence the triggering of events far in the future.
To quantify this slow decay, Table~\ref{tb:times} gives the values of 
two ``characteristic times'' $T_{0.95}$ and $T_{0.99}$ defined respectively by
$\int_0^{T_{0.95}} h(t)dt=0.95$ and $\int_0^{T_{0.99}} h(t)dt=0.99$. Intuitively, 
a given event occurring at some time $t_i$ has still a $5\%$ (resp. $1\%$) probability of triggering new events 
at times larger than $T_{0.95}$ (resp. $T_{0.99}$).
Taking the typical values $\tau_0=1$ second and $\epsilon=0.15$, 
we have $T_{0.95}=1.1\cdot10^{7}$ seconds and $T_{0.99}=1.3\cdot10^{9}$ seconds for approximate power law kernel~\eqref{eq:pow_jpb_sum},
which is about $488.8$ (resp. $57777.7$) trading days (where each trading day consists of $6.25$ hours). For the exact Omori-type kernel~\eqref{eq:pow}, these values are much larger:  $T_{0.95}=4.9\cdot10^{8}$ seconds and $T_{0.99}=2.3\cdot10^{13}$ seconds.
Even for the large windows of two months considered in \citep{HardimanBouchaud2013}, 
if one considers the power law kernels with small exponent $\epsilon$ as capturing the real
long memory of the empirical time series, this implies that
events that occurred years before a given window still exert a significant influence on the
occurrence of events in that window.
For numerical simulations, this implies that edge effects play a dominant role and may lead
to significant biases in parameter estimations, if not accounted for properly by using 
extremely long realizations and burning out a very long transient.

\newcolumntype{C}[1]{>{\centering}m{#1}}

\begin{table}[h!]
\caption{``Characteristic times'' $T_{0.95}$ and $T_{0.99}$ defined respectively by
$\int_0^{T_{0.95}} h(t)dt=0.95$ and $\int_0^{T_{0.99}} h(t)dt=0.99$ for the approximate power law kernel~\eqref{eq:pow_jpb_sum} and Omori-type kernel~\eqref{eq:pow}. }
\begin{center}
\begin{tabular}{C{1.5cm}C{2cm}cC{2cm}c}
\toprule
\multirow{2}{*}{$\epsilon~\Large/~\theta$} &  \multicolumn{2}{c}{Approximate PL kernel~\eqref{eq:pow_jpb_sum}} & 
 \multicolumn{2}{c}{Omori-type kernel~\eqref{eq:pow}} \\
& $T_{0.95}/\tau_0$ & ~~~$T_{0.99}/\tau_0$~~~ & $T_{0.95}/\tau_0$ & ~~~$T_{0.99}/\tau_0$~~~\\
\toprule
$0.1$ & $1.6\cdot10^8$ & $4.3\cdot10^9$ & $1\cdot10^{13}$ & $1.0\cdot10^{20}$ \\
$0.15$ & $1.1\cdot10^7$ & $1.3\cdot10^9$ & $4.9\cdot10^8$ & $2.3\cdot10^{13}$ \\
$0.2$ & $6.5\cdot10^5$ & $2.0\cdot10^8$ & $3.2\cdot10^6$ & $1.0\cdot10^{10}$ \\
$0.3$ & $9.2\cdot10^3$ & $1.6\cdot10^6$ & $2.3\cdot10^4$ & $4.8\cdot10^6$ \\
$0.5$ & $2\cdot10^2$ & $5\cdot10^3$ & $4.2\cdot10^2$ & $1\cdot10^4$ \\
$1.0$ & $13$ & $63$ & $20$ & $99$ \\
\bottomrule
\end{tabular}
\end{center}
\label{tb:times}
\end{table}

\begin{figure}[h!]
  \centering
  \includegraphics[width=0.9\textwidth]{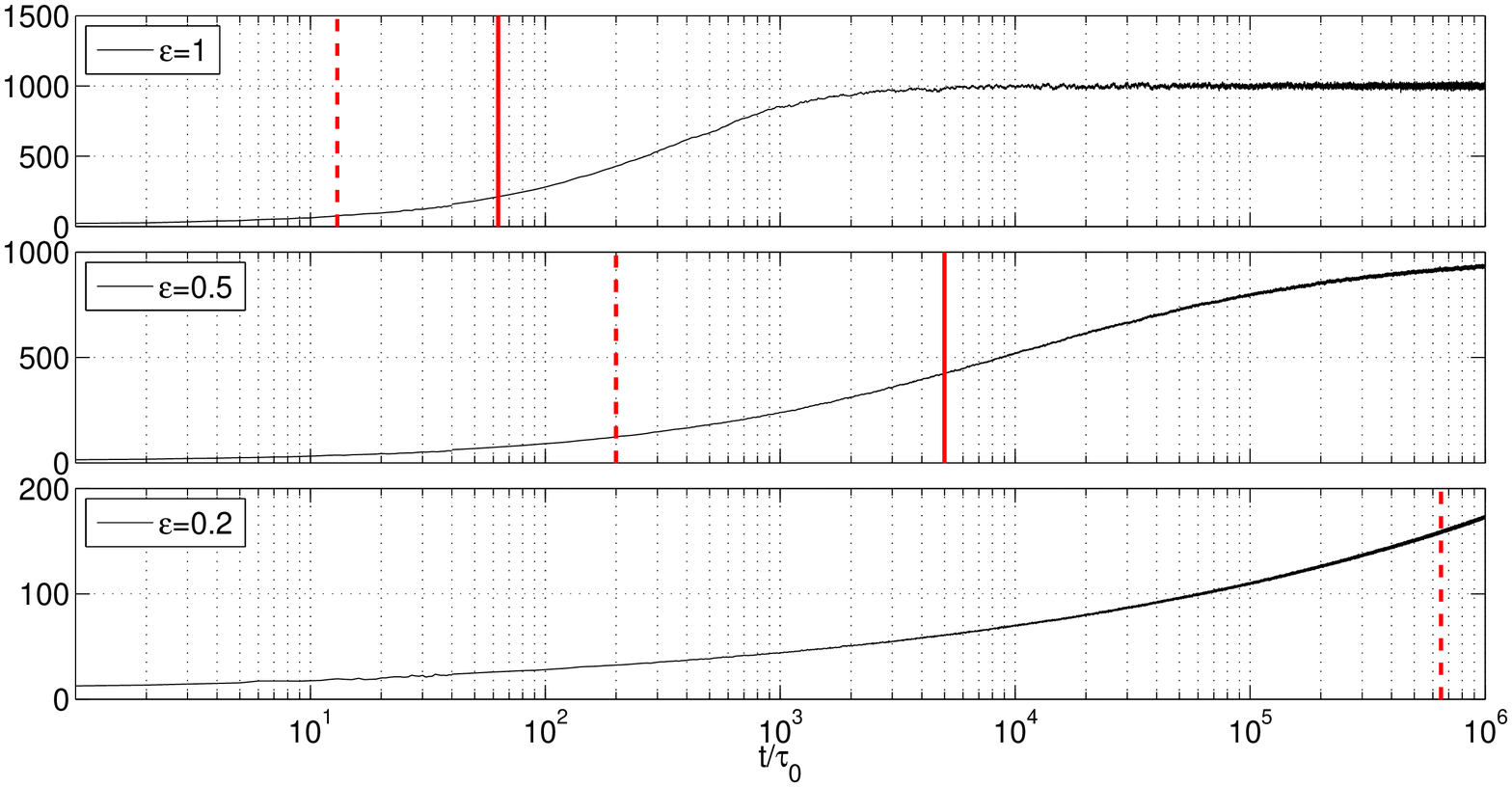}
  \caption{Dependence of the average number of events in 10 seconds intervals
as a function of time, where the origin of time $t=0$ is the start of the simulation with
no preceding ancestors. The simulations are performed
 for an almost critical ($n=0.99$) Hawkes process with approximate power law kernel~\eqref{eq:pow_jpb_sum} for $\mu=1$, $\tau_0=1$ and $\epsilon=1$ (upper panel), $\epsilon=0.5$ (middle panel) and $\epsilon=0.2$ (lower panel). Dashed and solid vertical lines corresponds to the values $T_{0.95}$ and $T_{0.99}$ respectively. The curves are obtained by averaging over 2000 realizations.} \label{fig:realization}
\end{figure}

The resulting edge effect is illustrated in Fig.~\ref{fig:realization}, which shows that 
the transient period, which is characterized by a strong non-stationarity, lasts much longer than 
the ``characteristic times'' $T_{0.95}$ and $T_{0.99}$. As a rule of thumb, 
one needs to wait at least 100 times more than $T_{0.99}$ 
in order to reach the ``quasi-stationary'' regime.

Such long transient periods obviously affect the estimation of the parameters, assuming that the underlying data is generated by the Hawkes process with long memory kernels, such as~\eqref{eq:pow}--\eqref{eq:pow_jpb_sum}. In order to illustrate this, we have simulated almost critical ($n=0.99$) Hawkes processes with the approximate power law kernel~\eqref{eq:pow_jpb_sum} in the very long interval $(0, 10^9+10^6]$. We have then ``burned'' 
a first interval of length $10^9$ seconds. As seen from Table~\ref{tb:times} and Fig.~\ref{fig:realization}, 
such a long interval is sufficient to avoid the transient dynamics for kernels with $\epsilon\ge0.5$.
But it is definitely not sufficient to get rid of edge effects for $\epsilon\le0.2$. 

\begin{figure}[h!]
  \centering
  \includegraphics[width=0.8\textwidth]{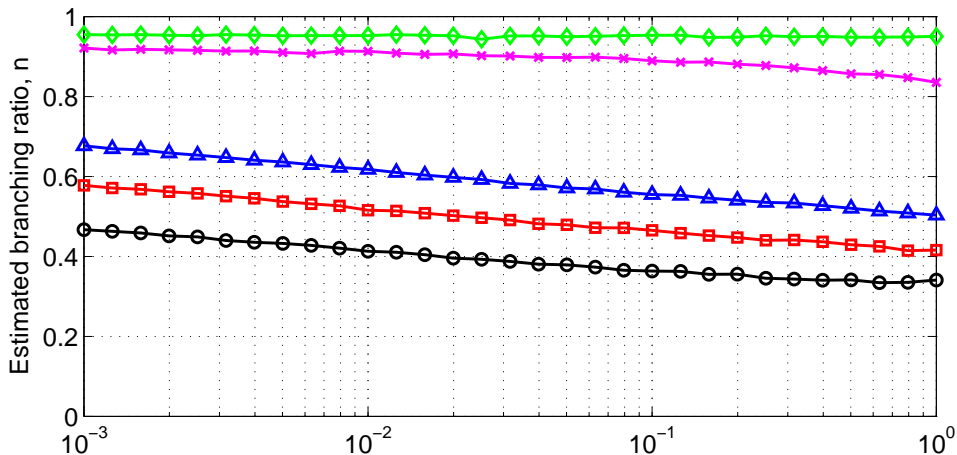}
  \caption{Results of the estimation of the branching ratio $n$ on synthetic time series 
  generated by the Hawkes process with the approximate power law kernel~\eqref{eq:pow_jpb_sum} with $\mu=0.02$ and $n=0.99$. The parameter $\tau_0$ is spanned in the interval $[10^{-3}, 1]$ (x-axis). Different lines correspond to different values of the exponent $\epsilon$: (bottom-up) $\epsilon=0.1$ (black circles), $\epsilon=0.15$ (red squares), $\epsilon=0.2$ (blue triangles), $\epsilon=0.5$ (magenta crosses) and $\epsilon=1$ (green diamonds). The figure corresponds to the case when the estimation is performed by using the Hawkes process with the exponential kernel~\eqref{eq:exp} and the whole interval $[0, 10^6]$ seconds is split into intervals of 30 minutes each for calibration. Solid lines represent averages over 555 estimations. The confidence intervals are extremely narrow and not presented (the maximum standard deviation of the estimation of $n$ is approximately $0.06$ for the upper panel and $0.013$ for the lower panel). } \label{fig:tau0}
\end{figure}

Our simulation experiment is aimed at determining possible biases in the estimation of 
the branching ratio $\hat n$ when the Hawkes model used for the estimation has a short-range exponential kernel~\eqref{eq:exp}, while the data is generated with the long-range generating kernel~\eqref{eq:pow_jpb_sum}. As described above, we have used the interval of $T=10^6$ seconds, which corresponds approximately to 2 months of trading (44 trading days of 6.25 hours each). We split the interval of $10^6$ seconds into subintervals of length $1800$ seconds (30 minutes --- in total 555 subintervals), however we did not apply any randomization to the timestamps (the effect of randomization will be considered in the following section). We then calibrated the Hawkes model with an exponential kernel~\eqref{eq:exp} in each of the $1800$ seconds intervals. 
The upper panel of Fig.~\ref{fig:tau0} shows that, notwithstanding the difference of the kernels
using in the generating and in the estimating Hawkes process, the calibration recovers
correctly the near critical values of the branching ratio $n\ge 0.9$ for $\epsilon\ge 0.5$, i.e. when the edge effect was successfully removed. However for $\epsilon\le0.2$, when the transient effect is strong and 
only a minor fraction of the ancestors fall within the window of analysis, the estimations present significant downward biases. The slow increase of the estimated branching ratio $\hat n$ as $\tau_0$ decreases 
reflects the corresponding decrease of the transient period (see Table~\ref{tb:times}).

Simulation of long time-series can be made computationally efficient. The traditional simulation method
of Hawkes self-excited processes uses the modified thinning procedure~\citep{Lewis1979,Ogata1981}, which is based on simulating the homogeneous Poisson process with sufficiently high intensity and then accepting or rejecting points with a probability given by the Hawkes model. The numerical complexity of this method is $\mathcal{O}(N^2)$, which prevents its use for simulating long time-series of near-critical Hawkes processes. A much more efficient simulation method
benefits from the branching structure representation of the Hawkes process discussed in section~\ref{sec:branching}~\citep{Moller2005,Moller2006}. One can construct the Hawkes process as a combination of the homogeneous Poisson process with intensity $\lambda(t)=\mu$ that describes immigration, and of a set of non-homogeneous Poisson processes with intensities $\lambda_i(t)=n h(t-t_i)$, which represent descendants. Using a vectorized construction of the ancestors for each generation of descendants, one can reach a numerical complexity of $\mathcal{O}(\mu T\cdot M)$, where $T$ is a window size and $M$ is the number of generations that fit this window.

\subsection{Multiple extrema of the likelihood function and suboptimal solutions}\label{sec:extrema}

The calibration of the Hawkes model requires 
finding the numerical solution of the minimization of $-\log L(\cdot)$ (negative of the log-likelihood function given
by expression \eqref{eq:loglik}) in the parameter space $\{\mu,n,c,\theta\}$ or $\{\mu,n,\tau_0,\epsilon\}$
(for the power law kernels). 
When the data is generated by the Hawkes model \eqref{eq:hawkes_discr} with the same memory kernel as
that used for the calibration, as can be ensured in synthetic cases, the 4-dimensional ``cost function'' $-\log L(\cdot)$ typically 
has one pronounced minimum that can be easily found with almost any local minimization algorithm. 
However, in some cases when the memory kernel used for the estimation differs from the kernel 
used to generate the data, the cost function may have a very flat valley similar to
that associated with the Rosenbrock function~\citep{Rosenbrock1960}, 
which requires adaptive numerical descent methods.
For real data, in the presence of noise and possible outliers, and the highly probable fact that the
generating process does not coincide exactly with the Hawkes process, the cost function is 
in general non-trivial with several local minima. As a consequence, 
solutions found by local optimization methods are highly sensitive to the starting point of the method. 
In this case, local descent methods are not sufficient and should be complemented 
with some metaheuristics~\citep{Talbi2009Metaheuristics}, 
the simplest of which is running local minimization methods starting from
multiple starting points scattered within the whole search space and choosing the best solution among them.
Nevertheless, there is typically no guarantee that the solution thus found is the best one.

In order to illustrate the problem associated with the existence of multiple local minima 
and its consequence, we visualize the cost function in a two-dimensional space, obtained
by partitioning the four-dimensional search space $\{\mu,n,\tau_0,\epsilon\}$ 
into subspaces $\{\mu, n\}$ and $\{\tau_0,\epsilon\}$ and subordinating the first to the second, as described in Section~\ref{numerical}. We consider the concrete example of the
E-mini S\&P 500 Futures Contracts with millisecond resolution, where periods of two months
are used for the calibration, in which only the Regular Trading Hours (9:30 to 16:15 CDT) 
are kept for each day, which are glued together in a single continuous time series.
In addition, a kind of detrending is performed to remove the 
intraday U-shape (``lunch effect'')  by using a function $w(t)$ that we estimated 
over the same two months interval (see section~\ref{sec:detrending}). Before calibration,
we additionally applied a randomization procedure to the timestamps within millisecond intervals\footnote{In the sections~\ref{sec:overnight}--\ref{sec:regime}, we discuss in details every step of this data preprocessing and show that all of them result in significant upward biases for the estimated branching ratio.}.
 
\begin{figure}[h!]
  \centering
  \includegraphics[width=0.8\textwidth]{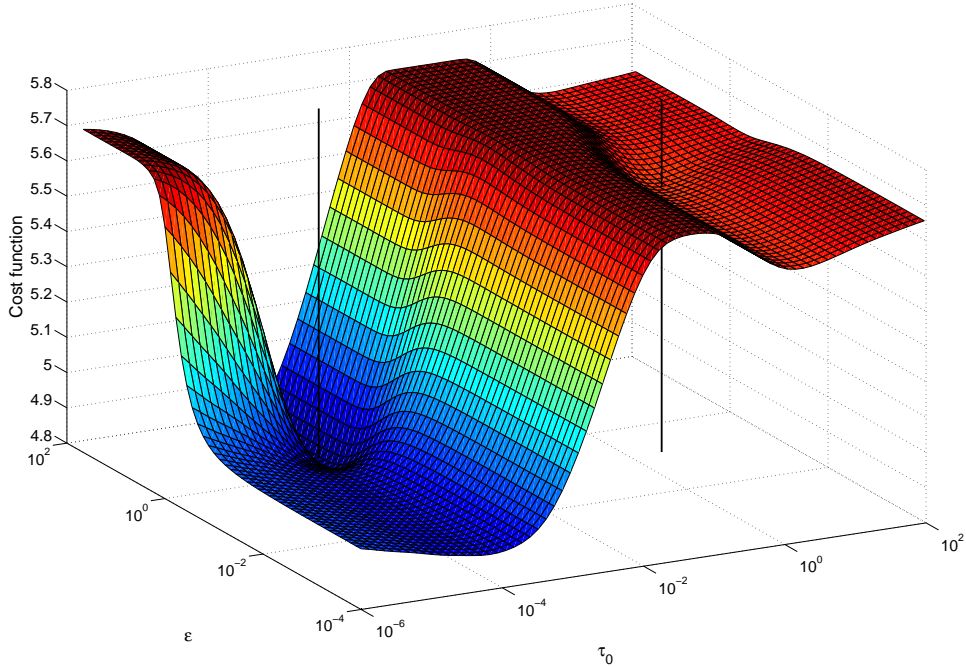}
  \caption{Surface of the cost function $S(\tau_0,\epsilon|t_1,\dots, t_N)$ (eq.~\ref{eq:optim_2}) 
  used in the calibration of the Hawkes model for the period March 1, 2001 --- April 30, 2001 (see text for details). 
  The two black vertical lines point to the locations of the two local minima, global on the left
  and local on the right.} \label{fig:surf}
\end{figure}

Figure~\ref{fig:surf} shows the typical surface of the cost function $S(\tau_0,\epsilon|t_1,\dots, t_N)$
for the period 1998--2005. Even the reduced cost function (\ref{eq:optim_2})
keeps a rather complex structure with several minima and very long valleys where the minimization algorithm 
can be stuck. The global minimum of the cost function (pointed out by the vertical black line 
on the left of fig.~\ref{fig:surf}) gives $\hat\mu=0.3031$, $\hat n=0.0751$, $\hat\tau_0=0.00028$, $\hat\epsilon=2.4604$ and the corresponding value of the cost function (negative log-likelihood) is $S=2.09\cdot10^5$. 
However, if the starting point for the minimization algorithm is chosen 
incorrectly (for instance, if the minimization is started from the point $\tau_0=1$, $\epsilon=1$), 
then the optimization procedure converges to the local minimum (vertical black line on the
right of fig.~\ref{fig:surf}), which gives $\hat\mu=0.0150$, $\hat n=1.1054$, $\hat\tau_0= 2.8089$, $\hat\epsilon=0.1442$ and a much higher value of the cost function $S=2.85\cdot10^5$. These two solutions 
belong to completely different regimes of the process: while the optimal set of parameters 
points to a subcritical regime ($\hat n=0.0751\ll1$), the sub-optimal solution 
diagnoses a super-critical regime ($\hat n=1.1054>1$). 

On the same dataset and using the same procedure, \cite{HardimanBouchaud2013}
have reported estimates for $n$, $\tau_0$ and $\epsilon$ that are very close to those 
corresponding to the local suboptimal minimum $\hat n=1.1054$, leading these authors to 
incorrectly conclude about the existence of criticality or super-criticality at all times. 

\begin{figure}[h!]
  \centering
  \includegraphics[width=0.8\textwidth]{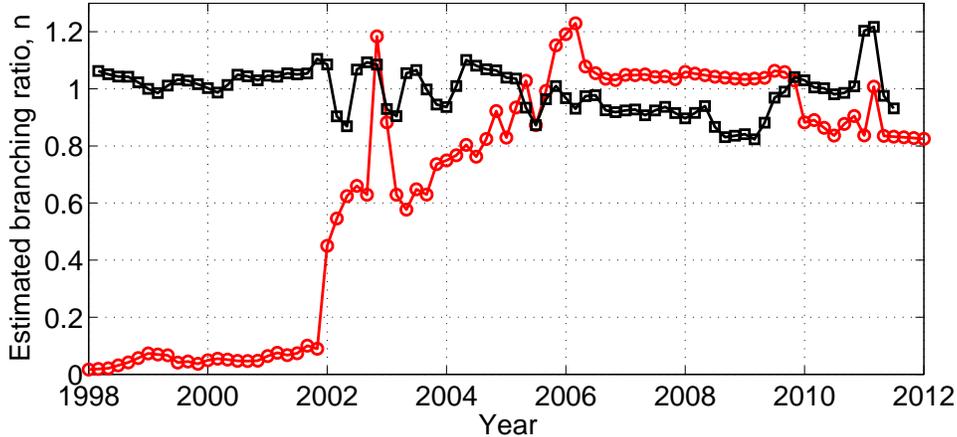}
  \caption{(Bottom) Estimate of the branching ratio for the global minimum (red circles) and ``suboptimal'' estimates presented in~\citep{HardimanBouchaud2013} (black squares).} \label{fig:cost}
\end{figure}

The difference between the local suboptimal and global optimal minimization of the cost function
is illustrated in Figure~\ref{fig:cost}.
This allows us to compare the ``suboptimal'' branching ratios reported 
in~\citep{HardimanBouchaud2013} with the ``optimal'' ratio obtained by 
exhaustive exploration of the parameter space. 
One can see that the estimates presented in~\citep{HardimanBouchaud2013} are far from being optimal and this leads to the incorrect conclusion that the system has been critical and super critical over most of time since 1998,
while the global optimal solution shows a low branching ratio until 2002, which grows
progressively until 2006 and then later fluctuates between $0.8$ and $1.1$ approximately.

Finally, we stress that the issue raised here is even more crucial for the multivariate Hawkes model. The monovariate Hawkes process with, for instance, exponential kernel~\eqref{eq:exp} and constant background activity is fully described with a set of only 3 parameters ($\mu$, $n$ and $\tau$). With the addition of one dimension and accounting for the cross-excitation (bi-variate model that is used, for instance, in~\citep{Bowsher2007,Toke2011,AitSahalia2011,Embrechts2011,Muzy2013hawkes}), the parameter set is increased to 10 parameters when no symmetry is imposed. The six-variate Hawkes model suggested in~\citep{Cont2011} is parametrized by 78 parameters in the general case. The 10-variate Hawkes model~\citep{Large2007} in general would require the calibration of 210 parameters (however, \citep{Large2007} reduces the number of calibrated parameters to 32). Obviously, 
the augmentation of the parameter set makes the cost function even more complicated. This increases
the chances of multiple extrema and thus poses a numerical challenge to the calibration procedure.

\section{``Microstructure'' of the high-frequency financial data and biases \label{sec:microstructure}}

The following four subsections discuss some properties of the high-frequency financial data that should 
be taken into account in any analysis using point process models. Since they have been been inadequately 
addressed in \citep{HardimanBouchaud2013} and are rarely mentioned
in the literature, it is useful to develop them for future use. 

Before proceeding we need to note, that high-frequency data cleaning is one of the most important aspects of any modeling, since presence of outliers or misrecorded data may substantially bias estimates and forecasts. Issues of filtering and cleaning the high-frequency will not be discussed here, and we refer reader to the  existing literature~\citep{Dacorogna2001_hifreq,Falkenberry2002,Brownlees2006,BarndorffNielsen2009}.

\subsection{Regular Trading Hours and overnight trading}\label{sec:overnight}

The first question concerns the choice of the period of analysis. In 
our original work~\citep{FilimonovSornette2012_Reflexivity}, we have calibrated the 
Hawkes process within short intervals of 10 to 30 minutes that we scanned over the whole history 
of E-mini S\&P 500 Futures Contracts. Then, we collected the estimates within 
Regular Trading Hours (RTH, 9:30 to 16:15 CDT) in each month and averaged them to construct 
the ``reflexivity index'' for the given month~\citep{FilimonovSornette2012_Reflexivity,FilimonovSornetteUNCTAD2012_Commodities}. 
In contrast, \cite{HardimanBouchaud2013} considered long intervals of two months, where 
data was concatenate from the Regular Trading Hours of each day to construct one single continuous point process. 

Choosing RTH is motivated by two important reasons. First, the trading activity within the RTH is typically larger than within the overnight period. Second, the trading activity within each 24 hours is highly non-stationary, where both intensity and dynamics within RTH and overnight periods differ significantly. When using short time intervals
of 10 to 30 minutes, there is no problem with focusing only on the RTH.
In contrast, throwing away a significant part of the day (16.25 hours versus 6.25 RTH) and 
concatenating data from day to day over a two-month period for the kind of long-term
analysis performed in \citep{HardimanBouchaud2013} may lead to significant and
uncontrollable biases.

\begin{figure}[h!]
  \centering
  \includegraphics[width=\textwidth]{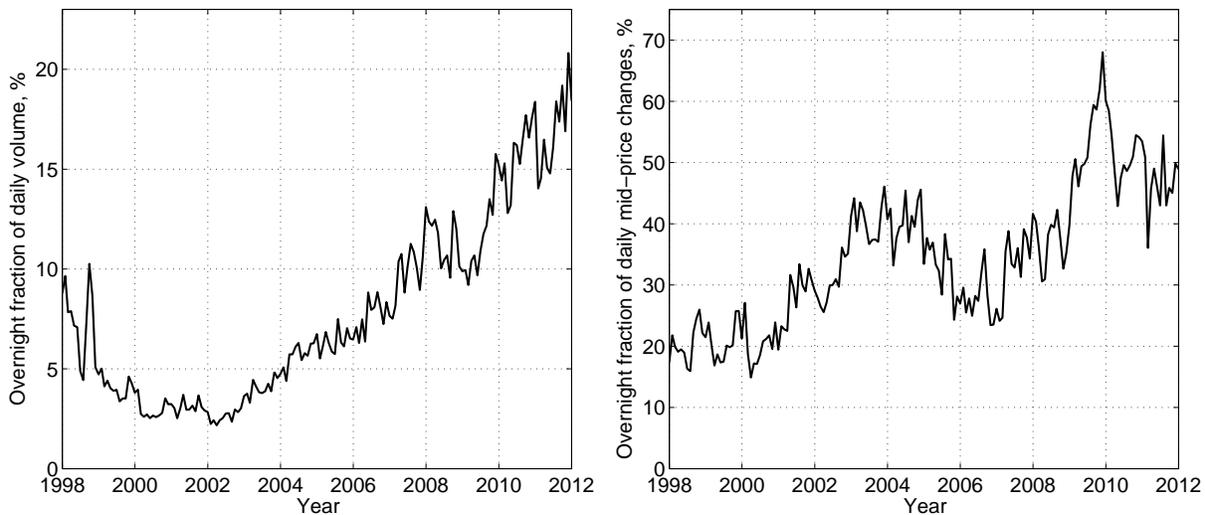}
  \caption{(Left) Fraction of total daily volume that is traded outside of Regular Trading Hours (9:30 to 16:15 CDT) on E-mini S\&P 500 Futures Contracts. (Right) Fraction of total daily number of mid-quote price changes that is observed outside Regular Trading Hours on E-mini S\&P 500 Futures Contracts. Each point represents an average value calculated over two months intervals.} \label{fig:overnight}
\end{figure}

The electronic trading of E-mini S\&P500 Futures is active 23 hours a day. And while the volume traded over night is lower than the volume traded in RTH, this difference is not that extreme. As seen from fig.~\ref{fig:overnight} (left), in recent years, almost 20\% of the daily volume is traded outside of the RTH. Moreover, the number of limit orders submitted over night and the number of mid-quote price changes triggered by them (which are the input of the analysis in~\citep{FilimonovSornette2012_Reflexivity,FilimonovSornetteUNCTAD2012_Commodities,HardimanBouchaud2013}) 
is huge. As seen in fig.~\ref{fig:overnight} (right), since 2002, the number of mid-quote price changes over night consistently amounts to more than 30--40\% of all mid-quote price changes and, in recent years, it reached the value of 50\%, and even 67\% in 2010. In other words, in 2010, the number of events (mid-quote price changes) observed over night 
was up to 2 times larger than the number of events observed in RTH and considered by~\cite{HardimanBouchaud2013}. 
Throwing away the over night  data is thus simply unwarranted, especially when using long memory power law kernels
that link the activity across days, as done in~\citep{HardimanBouchaud2013}. 

By considering only regular trading hours (RTH) and by constructing single continuous time series, one ignores a substantial part of the data. This data manipulation no doubt will affect the calibration of the model and in particular the estimates of the branching ratio. Indeed, the ignored periods might contain zero-order events that are ancestors of many 
of the events observed during RTH, which will result in 
an underestimation of the background activity and an overestimation of the branching ratio.
For example, during pre-market auctions the important information about corporate and macroeconomic events, 
which is usually released before trading hours, is being digested by the equity prices. Thus, not accounting for this process will overestimate the total reflexivity of all equity, futures and derivatives markets.
Or, on the other hand, if most of zero-order events are concentrated in RTH, the endogenous 
part of the intensity and the branching ratio will be underestimated. 
However, as we will see further, accounting for this non-stationarity is not trivial.

Moreover, practically speaking, there exists three substantially different trading periods for E-mini S\&P 500 Futures Contracts: (i) Regular Trading Hours of the US markets (9:00--16:00 EST), (ii) Regular Trading Hours of the Asian markets (20:00--3:00 EST) and (iii) Regular Trading Hours of the European markets (3:00--11:00 EST). Market participants and their strategies differ substantially over these trading periods. It is difficult to expect that a single model with fixed parameters can describe all three regimes simultaneously and arguably it is better to consider the three intervals (i)--(iii) separately. While treating the three intervals separately is feasible
within the approach of~\citep{FilimonovSornette2012_Reflexivity,FilimonovSornetteUNCTAD2012_Commodities} that considers short analysis windows to construct an effective monthly ``index'' out of these estimates, in contrast
the approach of Hardiman et at.~\cite{HardimanBouchaud2013} is inapplicable here, as it requires 
the construction of long continuous time series.

\subsection{Latency, grouping of timestamps and the ``bundling effect''}\label{sec:millisecond}  

Despite the fact that many high frequency data providers give millisecond (or even sometimes microsecond) 
precision for tick timestamps, a rather large number of transactions and quote changes have identical timestamps. The origin of this phenomenon lies in the nature of the data feed from the exchange, which is obtained by the FAST/FIX protocol, which is nowadays the most commonly used protocol for communicating financial data. The protocol bundles multiple updates of multiple instruments within a single message by an algorithm designed by the exchange. Then, the package is sent to the data provider collection system, and which stamps the time of events on the package arrival, but not to the time when the transactions (quotes) were actually executed and recorded by the exchange. The exchange time, which is the only reliable timestamp, is coded in the FAST/FIX protocol and stamped with a resolution of seconds due to the protocol limitation. The (millisecond) timestamp of the data provider enriches the second resolution, however it is subjected to an uncertainty, which arrises from
\begin{enumerate}[(i)]
\item overhead brought by processing data on both sides;
\item latency of the message traveling from the exchange to data provider's collection system;
\item grouping multiple events into single FAST/FIX packages.
\end{enumerate}
Data processing (i) include time for packing/unpacking FAST/FIX package, which is of order of several tens of microseconds~\citep{Lockwood2012_FPGA}, and  overhead brought by operating with large-scale databases, which vary from milliseconds to hundreds of milliseconds or even seconds, depending on the implementation and scale.
The latency to the exchange (ii) is usually of the order of tens of milliseconds. These two factors introduce an effective bias to the timestamps, which in most cases is more-or-less regular and could be considered to be constant. 
Therefore, it is not usually relevant in the calibration of the Hawkes process, which is invariant 
with respect to a time shift. However, as we will see further, these factors can fluctuate a lot, which may affect the calibration of the parameters.

In most cases, the strongest bias to the timestamps is introduced by the bundling of events into FAST/FIX packages (iii), which results in the fact that the the actual time of any tick is uncertain within a range that is larger than or equal to the time between two consecutive FAST/FIX packages. In contrast to (i) and (ii), this uncertainty is both irregular (for any particular event) and has much higher scales. For most exchanges, this time varies from tens of milliseconds in recent years to several hundreds of milliseconds or even seconds in early 1998--2003~\citep{FilimonovSornetteUNCTAD2012_Commodities}. 

\begin{figure}[h!]
  \centering
  \includegraphics[width=0.9\textwidth]{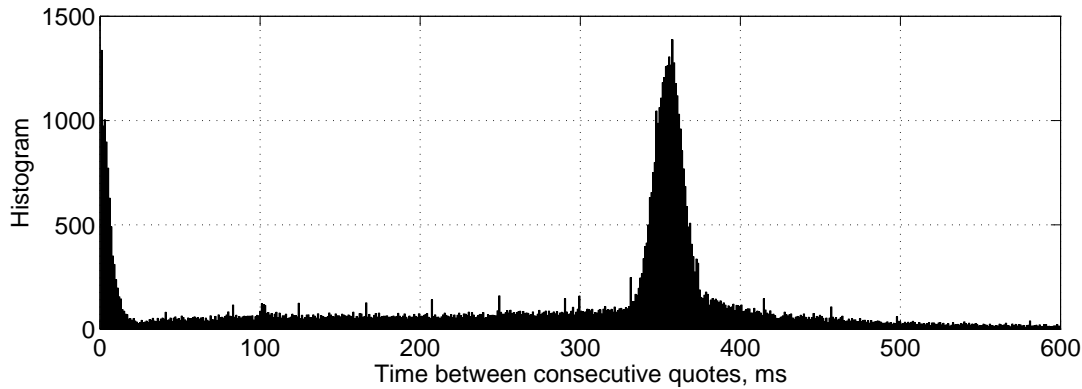}
  \caption{Histogram of the non-zero inter-quote time for E-mini S\&P 500 Futures Contracts on October 2, 2002 over Regular Trading Hours. Data source: Thompson Reuters Tick History.} \label{fig:hist_quotes_2002}
\end{figure}

Figure~\ref{fig:hist_quotes_2002} illustrates the irregularity in the time between consecutive quotes, recorded by the Thompson Reuters Tick History (TRTH). One can observe a strong peak of the histogram at the inter-quote time of approximately 350 ms, which reveals the high irregularity of the data on the arrival times due to the mechanisms described above (bundling of multiple events in packages and quasi-regular delivery of these packages from the exchange to data collection system). Naively, one could assume that this peak at 350 ms marks the time between deliveries of consecutive FAST/FIX packages. However, this assumption is very far from being correct.

In order to ensure data integrity, each FAST/FIX package that is sent from the exchange carries
a unique (within a trading day) sequential message number. Some data providers (such as Thompson Reuters Tick History -- TRTH) record these numbers along with the information about the quote or trade itself, which allow us to retrospectively evaluate the uncertainty of timestamps of quotes (or trades). The time between FAST/FIX packages can be estimated as the difference in timestamps of the first recorded quotes in successive packages. Another important characteristic that we analyze here is the overhead for data processing. This is particularly important for the TRTH and similar data provider, which timestamp events not at the moment when the FAST/FIX packet arrives to the collecting system but at the moment when the actual trade or quote is written to the database. The procedure of the data processing in TRTH is the following: the collection system receives packages from many exchanges containing updates for many instruments. All these packages are uncompressed and events (trades and quotes) are dispatched to the queue, from where they are written to the database. The additional uncertainty due to this pipeline might be significant. In order to estimate this data processing overhead, we considered the difference in timestamps between the last and the first quotes with identical sequential package number (i.e. events within the same package).

\begin{figure}[h!]
  \centering
  \includegraphics[width=\textwidth]{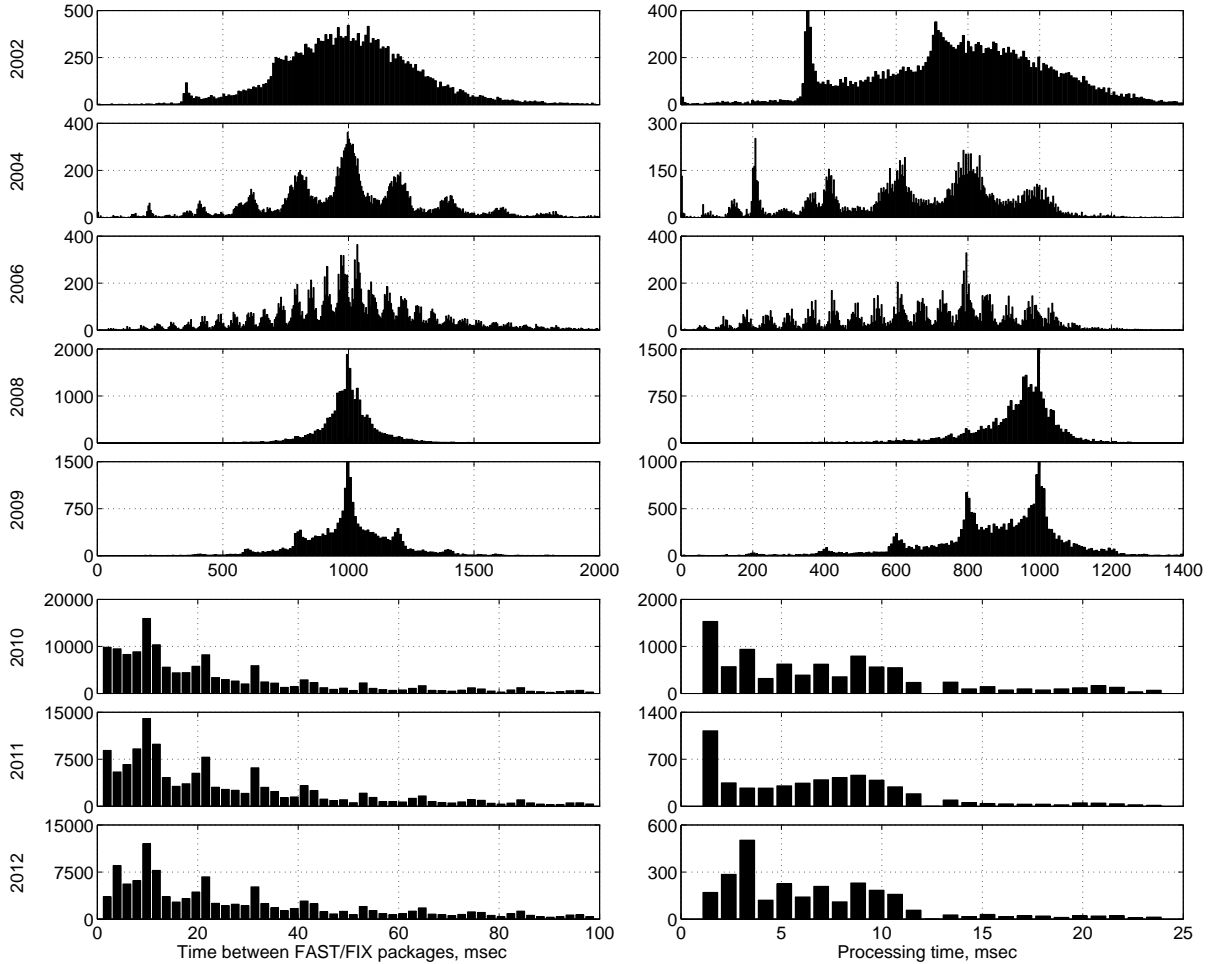}
  \caption{Histograms of the time between consecutive FAST/FIX packages (left panels) and overhead for the data  processing (right panels) for E-mini S\&P 500 Futures Contracts over Regular Trading Hours on different dates: (top--bottom) October 02, 2002; October 08, 2004; October 10, 2006; October 10, 2008; February 12, 2009; October 08, 2010; October 10, 2011 and October 10, 2012. Data source: Thompson Reuters Tick History.} \label{fig:hist_fastfix}
\end{figure}

Such an analysis is illustrated in Fig.~\ref{fig:hist_quotes_2002}, which uncovers
that the inter-package times are much larger than naively estimated above 
at about 350 ms (Fig.~\ref{fig:hist_fastfix}, upper panels). We observe that the time between consecutive FAST/FIX packages for the same date vary from 300 to 1700 milliseconds, having an average of approximately 1 second. Moreover, the processing times is also very large --- one package could take up to 1300 milliseconds. 
In a worst case scenario, this can sum up to 3 seconds. Note that this effective resolution 
can be significantly lower than the resolution of timestamps from the exchanges (1 second). Then,
the peak in Fig.~\ref{fig:hist_quotes_2002} can be simply explained as corresponding, not to
the inter-package time but, to the time difference between the end of processing a previous package and 
the start of processing the next one. We verified this explanation directly by analyzing the corresponding histogram.

Fig.~\ref{fig:hist_fastfix} presents the histograms of waiting times between packets and of processing times in different years for the TRTH database. Before 2009, both sources of uncertainties in the timestamps were enormous, having scales of 500--1000 milliseconds. In the second half of 2009, the data quality increased significantly
and these uncertainties dropped to values of tens of milliseconds. Note however
that these uncertainties of tens of milliseconds are still larger than the 
often claimed ``millisecond resolution''. 

The frequency of package arrivals depends on multiple factors. First, it obviously depend on the protocol itself: The FAST (FIX Adapted for Streaming) protocol was introduced in 2006 and had two revisions in 2007 (revision 1.1) and 2009 (revision 1.2). The underlying FIX (Financial Information eXchange) protocol also passed through several 
evolutionary steps, from its introduction in 1992 to the modern versions FIX~4.3 (2001/2002), FIX~4.4 (2003) and FIX~5.0 (2006/2008/2009).
The adoption time of protocols is different in different exchanges, and each exchange may implement some specific rules for collecting messages. However, for the same exchange, all customers with similar subscription types will receive quote updates with the same frequency. In contrast, additional overhead for data processing may vary from one data provider to another and depend on the design of the collection system.
Therefore, our estimates are indeed conservative in the sense that latency and waiting time depend on the infrastructure of the collecting system and the type of data-feed subscription. In practice, one can expect that the exchange gateway can send up to tens of thousands of FAST/FIX messages per second. However such throughput is available for co-located collecting system.

Two possibilities can be considered for dealing with the uncertainty in the timestamps resulting from the FAST/FIX protocol. One is to consider only the timestamps provided by the exchange (with resolution of seconds) as a reliable source of data. The other is to use the enriched millisecond timestamps of TRTH, while accounting for the uncertainty due to bundling updates in FAST/FIX packages (as for instance in~\citep{FilimonovSornetteUNCTAD2012_Commodities}). In any case, it is unacceptable to ignore the uncertainty in the timestamps by assuming a millisecond resolution of timestamps.

In order to account for timestamp uncertainties in the calibration of the Hawkes process,
we proposed a randomization procedure, which consists in uniformly redistributing timestamps within the intervals of uncertainty~\citep{FilimonovSornette2012_Reflexivity}. Essentially, this amounts to assuming that each event occurring within the interval of uncertainty is independent of all the others within the same interval (but not between different intervals). This raises the important question of what will happen if the interval, in which the randomization is employed, is different from the real time interval in which events are bundled to packages. In order to illustrate the possible bias that could stem from an improper assumption, we develop the following numerical simulations.

\begin{figure}[h!]
  \centering
  \includegraphics[width=\textwidth]{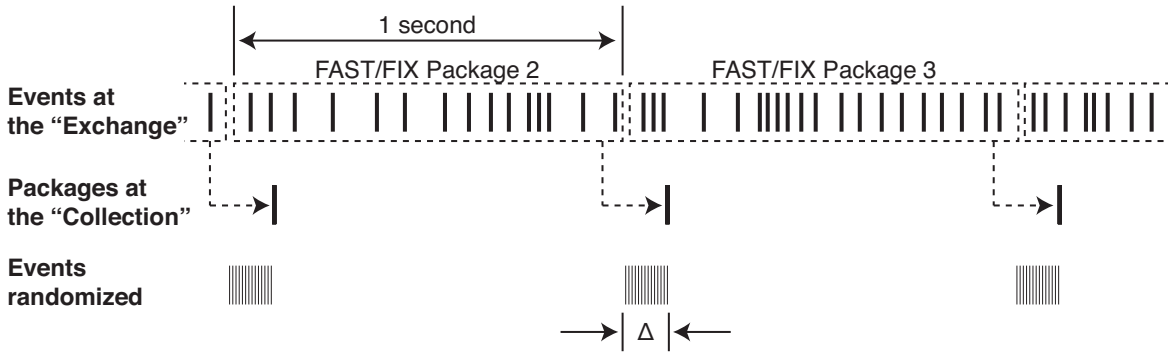}
  \caption{Illustration of the packaging of events at the exchange and the randomization procedure (see text).} \label{fig:events}
\end{figure}

We assume that there is no overhead for data processing (or it is constant) and all events are bundled by the exchange in packages every second and transmitted to the data collection system. However, for the calibration, we randomize the timestamps within a smaller intervals $\Delta$ (e.g. $\Delta=1$ millisecond as in~\cite{HardimanBouchaud2013}). Obviously, this will result in spurious clustering of events (Fig.~\ref{fig:events}) that will be reflected in the estimation of the branching ratio. It turns out that the resulting overestimation of the branching ratio is similar to the effect of the outliers that we considered in section~\ref{sec:robustness}.

For illustration of the possible resulting bias, we have simulated the Hawkes process with $\mu=3.5$, $n=0.5$ and (i) exponential kernel~\eqref{eq:exp} with $\tau=0.6$; and (ii) approximate power law kernel~\eqref{eq:pow_jpb_sum} with $\epsilon=1$, $\tau_0=0.3$. We have also considered (iii) a Poisson process with intensity $\lambda_0=7$ as a generating model. All the parameters are chosen to mimic the distribution of the number of events per second observed for E-mini S\&P 500 Futures Contracts on October 2, 2002 (Poisson-like distribution with mean $6.8$, median $7$ and standard deviation $3.7$). We ``grouped'' events into packages over $1$ second intervals and randomized them in intervals of duration $\Delta$, where $\Delta$ is varied from 1 millisecond to 1 second. Then, we calibrated these data
with the Hawkes model with the exponential kernel in case (i) and with the approximate 
power law kernel in cases (ii) and (iii).

\begin{figure}[h!]
  \centering
  \includegraphics[width=0.8\textwidth]{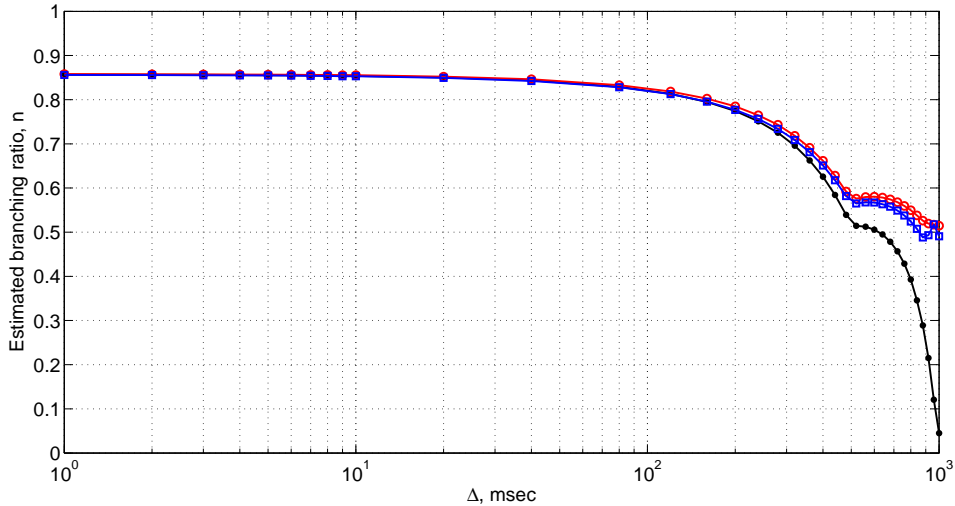}
  \caption{Illustration of the bias that results from improper assumptions on the duration  $\Delta$ of randomization intervals (see text) when the generating model is (i) the Hawkes process with an exponential kernel (red line with circles), (ii) the Hawkes process with the approximate power law kernel (blue line with squares) and (iii) a Poisson process (black line with dots). Each point represents an average over 100 realizations of length $T=10^5$ seconds (the initial interval of length $10^6$ seconds was ``burned'').  The confidence intervals are extremely narrow and not presented (the maximum standard deviation of the estimation of $n$ is $0.04$).} \label{fig:msec_sim}
\end{figure}

\begin{figure}[h!]
  \centering
  \includegraphics[width=0.7\textwidth]{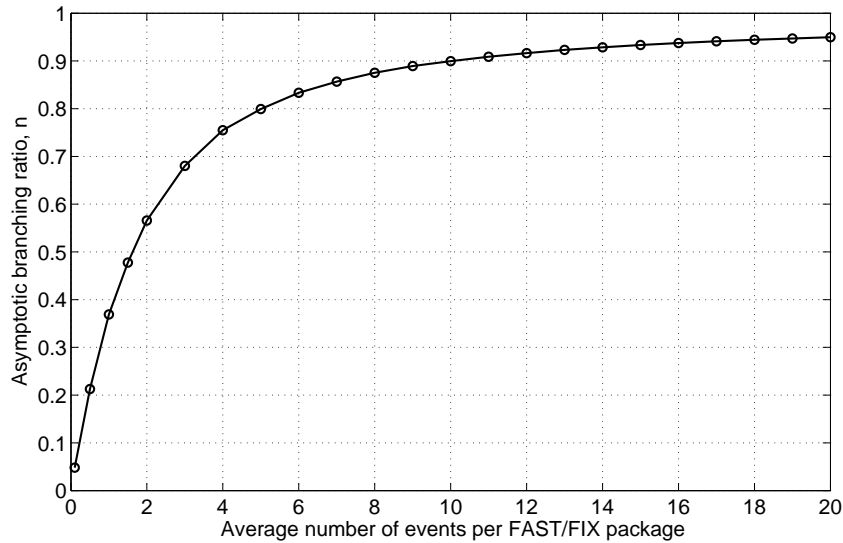}
  \caption{Dependence of the asymptotic value of the estimated branching ratio $\hat n$ when 
  the randomization interval duration $\Delta \to 0$
  (see text). The graph is obtained by numerical simulations of the homogeneous Poisson process with different 
  intensities $\lambda_0$, randomizing the data as described in the text and by calibration of the Hawkes model with exponential kernel~\eqref{eq:exp} on the Poisson generated data.} \label{fig:n_asympt}
\end{figure}

Fig.~\ref{fig:msec_sim} shows the dependence with $\Delta$ of the estimated branching ratio $\hat n$ in all three cases. 
We see that the estimated $\hat n$ monotonously increases when decreasing $\Delta$, having a cusp at $\Delta=0.5$ seconds, i.e., when the randomization interval becomes smaller than one half of the original ``bundling'' interval. For
smaller values of $\Delta$, the estimated branching ratio reaches an asymptotic value, which
is independent of the kernel used for the estimation of the branching ratio as well as of the 
underlying generating model. Indeed, the Poisson model, represented 
by the black line in Fig.~\ref{fig:msec_sim}, has the same spurious value $\hat n\approx 0.856$, 
as does the subcritical Hawkes model with $n=0.5$. This spurious value is only dependent on
the average number of events per FAST/FIX package, which are aggregated over 1 second interval in this case. 
Fig.~\ref{fig:n_asympt} illustrates this dependence, from which one can conclude that, in the worst case scenario, 
one can observe near-critical values of the branching ratio when, in fact, the underlying model is pure Poisson
with independent events ($n=0$). This observation strongly suggests that any sound analysis should either 
confine itself to using only second resolution timestamps of the exchange (as in~\citep{FilimonovSornette2012_Reflexivity}) or be complemented with an extensive robustness analysis implying model calibration with different assumptions on $\Delta$ (as, for instance, performed in~\citep{FilimonovSornetteUNCTAD2012_Commodities}).


Interestingly, in the randomization procedure discussed above, the overhead for the data processing (when present) may mitigate the negative impact of bundling events into packages. Indeed, instead of being recorded at exactly the same time with periods of roughly 1 second (Fig.~\ref{fig:hist_fastfix}, left panels) which leads to spurious clustering when being randomized over 1 millisecond intervals, the events of the same package are being more or less uniformly distributed across most of the interval between two consecutive packages. Thus, only a fraction of 
the inter-event intervals (corresponding to the peak in Fig.~\ref{fig:hist_quotes_2002}) are affected 
by the incorrect procedure. In other words, for the TRTH database, these two biases 
present in the data somehow partially compensate each other in the
calibration of the Hawkes model. However, this compensation is not necessarily present for other data providers
that stamp events at their times of arrival to the collection system.

Finally, we draw attention to the fact that the findings of the present section 
challenge all studies concerned with inter-event times, such as~\citep{Gourieroux2002,JiangZhou2008,Scalas2005,Oswiecimka2005,Eisler2006,JiangZhou2009,Ivanov2004,Politi2008} and others, where the observations of heavy tailed distributions, long memory correlation functions and multifractal scaling of the inter-event durations may be strongly biased by the bursty nature of the raw data, subjected to the ``bundling'' effects.

\subsection{Intraday seasonality and detrending of the data}\label{sec:detrending}

As already mentioned, for a constant background intensity ($\mu(t)=\mu=\mbox{const}$) and 
in the stationary case $n<1$, the branching ratio $n$ is exactly equal to the fraction of 
the average number of endogenously generated events among all events 
\citep{Sornette2003geo,FilimonovSornette2012_Reflexivity}. We can therefore interpret $n$
as a ``reflexivity index'', which quantifies the degree of endogeneity or reflexivity
of the financial market \citep{soros_alchemy1988}.
When $\mu(t)$ is time varying, a correct estimation of $n$ is made much more difficult by the fact
the we do not observe $\mu(t)$ independently but only the sum of the exogenous component $\mu(t)$ 
(first term in the r.h.s. of expression (\ref{eq:hawkes_general}))
and of the triggered events (second integral term in the r.h.s. of expression (\ref{eq:hawkes_general})).
If the dynamics of $\mu(t)$ is convoluted, many parameters may be needed to capture its 
complexity, which makes the estimation less robust and often degenerate (several solutions of 
very different parameter values compete at the same level of the likelihood function).
As a consequence, the determination of the branching ratio $n$ may be severely hindered
and biased. 

The exogenous component  $\mu(t)$ is known unfortunately to exhibit complicated structures, 
such as the ``U-shape'' intraday seasonality, reflecting the fact that the trading activity 
varies significantly during a typical day, being large at the open, low at lunch time and larger
towards the close. To address this problem, one needs to impose some knowledge on $\mu(t)$,
either coming from some expert advice, a priori information or empirical treatment.
Once a model $\mu_M(t)$ is chosen, 
each day activity can then be corrected by using this $\mu_M(t)$ in order to obtain
an effective activity with an assumed constant effective $\mu$. The hope is to remove
in this way the U-shape intraday seasonality and other patterns. This detrending method to apply this correction
uses the same time-transform theorem utilized for the residual analysis in Eq.~\eqref{eq:xi}. 
Namely, if $\hat \lambda(t)$ is the intensity of the generating process for $\{t_i\}$, 
then the integral~\eqref{eq:xi} transforms the process $\{t_i\}$ into a 
stationary Poisson process $\{\xi_i\}$ with constant intensity $1$. Similarly,
one can assume that, if $\mu_M(t)$ is the intensity of the background process, then 
the integral $\xi_i=\int_0^{t_i}\mu_M(t)$ will transform the Hawkes generated time series $\{t_i\}$ 
with the time-varying background intensity into the Hawkes process $\{\xi_i\}$ 
with constant background intensity (which could even be assumed equal to $1$
if the estimation of $\mu_M(t)$ is faithful, thus further reducing the number of unknown parameters). 
While intuitive, note that this transformation is only an approximation because the 
time series $\{t_i\}$ and $\{\xi_i\}$ cannot be both generated by a Hawkes model.
This results from the fact that the time transformation  $\xi_i=\int_0^{t_i}\mu_M(t)$
distorts the kernel of the process: if, for instance, all events $\{t_i\}$ have the same memory function 
$\varphi(t-t_i)$, the events $\{\xi_i\}$ will have memory functions $\tilde\varphi(\xi-\xi_i;t_i;\mu_M(t))$ 
that depends both on the background intensity and on the original event time $t_i$. 
We expect this approach to become less and less
reliable as $n$ is closer to $1$. We find that, when $\mu(t)$ varies slowly and weakly, the detrending
provides reasonable estimates for $n$. However, it does not perform when in the presence
of strong non-stationarity. Similar results are obtained by just dividing each daily activity by
the model exogenous activity $\mu_M(t)$. Fundamentally, this results from the fact that
no detrending method can fully disentangle the exogenous from the endogenous events
in the presence of multiple generations of triggered events, other than by reconstructing
the full sets of genealogical trees \citep{Zhuang2004,Marsan2008,VeenSchoenberg2008,LewisMohler2011}, 
which is in general extremely difficult for time-varying $\mu(t)$. A misspecification of $\mu(t)$ 
clearly leads to biases in the estimation of the branching ratio $n$: if one underestimates
the true background intensity $\mu(t)$ at some times, exogenous events will then be 
attributed to the self-excitation component of the intensity, thus leading to an 
over-estimation of $n$; in contrast, choosing $\mu(t)$ as being similar to the total unconditional intensity
will lead to interpret most endogenous events as exogenous and will push $n$ close to $0$.

\begin{figure}[h!]
  \centering
  \includegraphics[width=\textwidth]{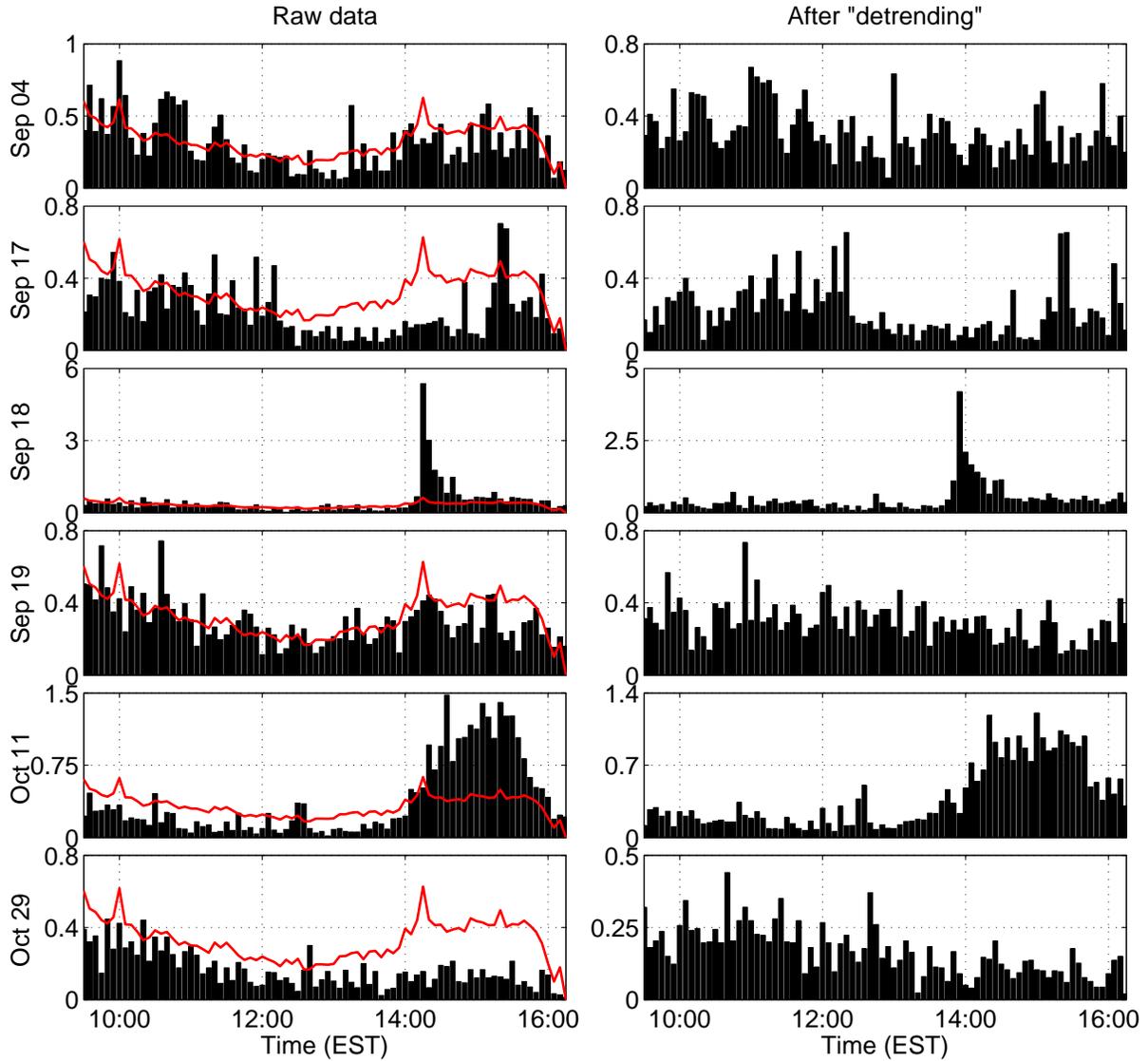}
  \caption{Estimated in non-overlapping 5-minute intervals unconditional intensity (in events-per-second) of flow of mid-quote price changes of E-mini S\&P 500 Futures Contracts on some dates of September--October, 2007. Left panels present the raw data (black bars) and the average intensity over the period of September~1--October~30, 2007 (red line). Right panels present the unconditional intensity after ``detrending'' using the average intensity.} \label{fig:ushape}
\end{figure}

Let us illustrate concretely these problems by using a standard approach, 
as done in~\citep{HardimanBouchaud2013}, which superimposes the trading activity of each 
day by matching the intraday time and taking the average over a large set of days
for each intraday time. This leads to define an ``average'' intensity $w(t)$
over a given time period  (e.g. a day) and assume that the background intensity 
follows this path, while particular features of each trading day are averaged out. 
In \citep{HardimanBouchaud2013}, $w(t)$ was assumed to be
a perfect periodic function with a period of one day, over the whole two-months interval of 
the analysis. Unfortunately, this assumption seems to be quite far from reality.
The left panels of Fig.~\ref{fig:ushape} illustrate that the assumed U-shape daily seasonality holds only on 
average, while every particular day has its own large scale deviation from this trend, i.e.,
the intraday activity cannot be viewed as just noise decorating the average assumed
average intraday pattern.
Moreover, for some days (such as September 17, 18 or October 11,  29 in Fig.~\ref{fig:ushape}),
the dynamics of trading activity does not follow a U-shape at all, having no significant drop
over lunch time. As a result, the detrending method using an average unconditional intensity does not fully remove the non-stationarity in the time series, as illustrated by the right panels of the Fig.~\ref{fig:hist_fastfix}.

A severe problem inherent on such seasonality detrending is that, by construction, it is supposed
that all deviations from the average trading activity are the result of endogenous self-excitations in the system.
The case of September 18, 2007 shown in Fig.~\ref{fig:ushape} is
an excellent illustration that this assumption may often be incorrect. At around 
14:00 EST, one can observe a large spike of trading activity that was accompanied almost immediately
by a spike in price (not shown in the picture). In fact, this is a direct consequence of the announcement of the Federal Interest rate by the FOMC commission, which surprised investors: the 
announced rate ($4.75\%$) was lower than the market expectations ($5\%$ according to the survey of Bloomberg~L.P.). In other words, the price jump and the spike in volatility are unambiguous exogenous event that resulted 
from the release of an unpredicted piece of information. In contrast, the average intensity approach
used in \citep{HardimanBouchaud2013} amounts to count it in significant part as being
endogenous. Note also that this single event September 18, 2007 weights in significantly in
the determination of $w(t)$, as can be seen by comparing the different scales in the ordinates
of the panels on the left, where the spike at around 14:00 EST is not really representative of the
trading activities of the other days while it constitutes a prominent feature of $w(t)$ due to 
the impact of the large activity on September 18, 2007. In fact, the release of almost all macroeconomic news
results in similar patterns of trading activity --- spikes at release times often preceded by a ``freezing'' of 
trading and a vanishing liquidity from the order book right before the announcement (see for instance \citep{Almgren2012}).

Informed by synthetic simulations, we will see in the following section that even small regime shifts 
lead to large overestimations of the branching ratio, if it is not accounted for in the model of the background activity. 
For this reason, spikes, such as the one that occurred in September 18, 2007, thus tend to drive the estimated branching ratio upward.  In order to avoid such non-stationarity issues in our analysis of short-term reflexivity~\citep{FilimonovSornette2012_Reflexivity,FilimonovSornetteUNCTAD2012_Commodities}, we have ignored all days with scheduled macroeconomic announcements, such as the FOMC rate decision 
or, in the case of oil futures, the US Energy Information Administration weekly report.

\subsection{Non-stationarity, regime shifts and mixing}\label{sec:regime}

\begin{figure}[h!]
  \centering
  \includegraphics[width=\textwidth]{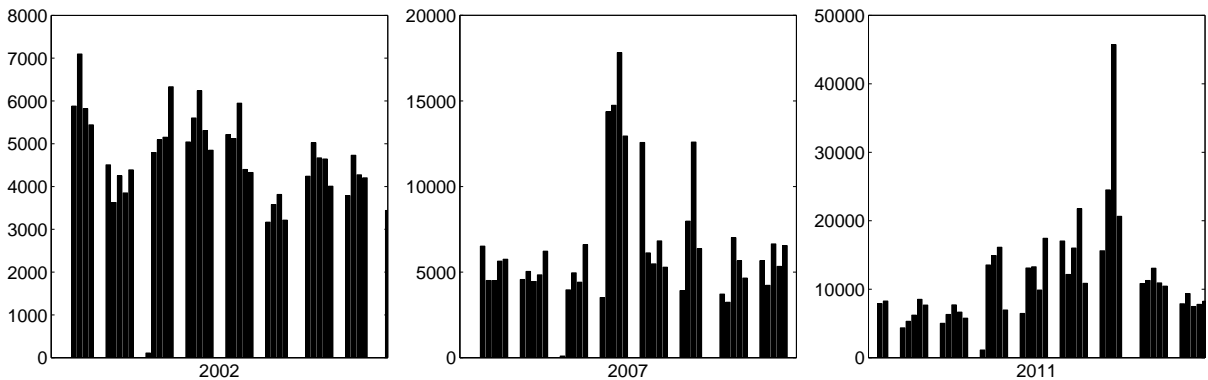}
  \caption{Dynamics of daily numbers of mid-quote price changes counted over Regular Trading Hours
 for the Front Month Contract of the E-mini S\&P 500 Futures from February 1 to April 1 in three different years.} \label{fig:regimes}
\end{figure}

Stationarity is extremely important in statistics. The Hawkes model is no exception and, in fact,
this model is very sensitive to the presence of any non-stationarity present in the time series. 
As discussed in section~\ref{sec:detrending}, intraday trading is subjected to the ``U-shape'' activity
pattern and spikes of activity on release of news. However, there is another perhaps even more important source of non-stationarity
in the fact that trading activity fluctuates significantly from day to day. 
Fig.~\ref{fig:regimes} shows the variations of the number of mid-quote price changes 
within a two month window in three different years.
One can observe that the activity fluctuates strongly within a factor 2 to 5 
in each such two month windows. Therefore, constructing a continuous time series 
in two month windows out of such 
non-stationary data would mix up different regimes, which again would seriously affect the calibration of the model.

We investigate the influence of such non-stationarity of the activity with synthetic time series.
Consider two independent realizations of the Hawkes process with identical kernels and identical parameters, 
but with different branching ratios $n_1$ and $n_2$. If we concatenate them and calibrate the Hawkes model
on the resulting time series, we will not estimate a branching ratio $\hat n$ for the combined time series
that is somewhere between $n_1$ and $n_2$ as we might expect intuitively. The regime switch will be 
interpreted by the Hawkes model as an excess clustering. Therefore, the estimated $\hat n$, which is also an effective measure of clustering for the Hawkes model, will tend to be higher so as to 
account for the regime switch. In fact, we find $\hat n$ always to be no less than the highest branching ratio of the individual sets.

\begin{figure}[h!]
  \centering
  \includegraphics[width=0.485\textwidth]{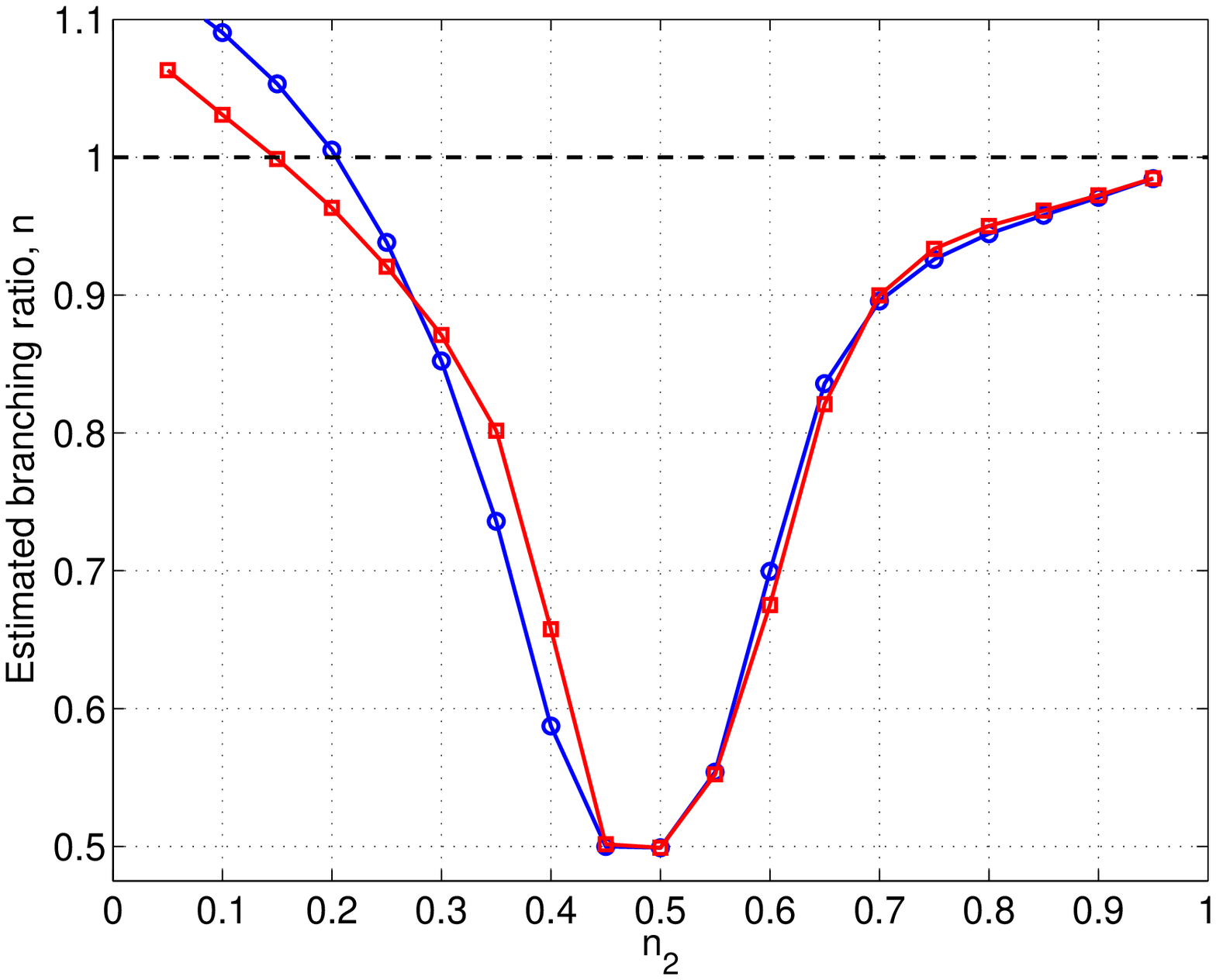}
  \includegraphics[width=0.485\textwidth]{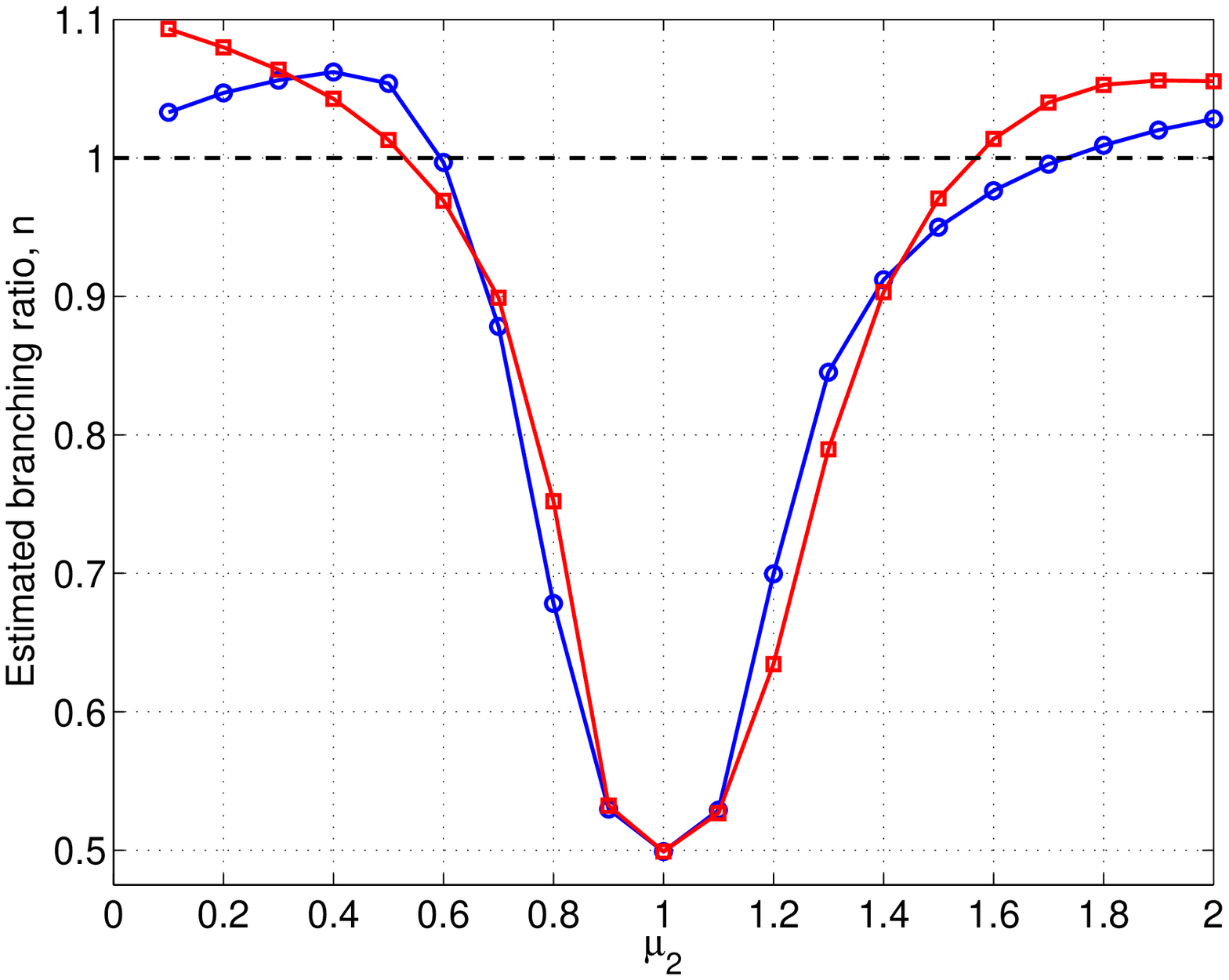}
  \caption{Results of the estimation of the branching ratio $\hat n$ on the synthetic time-series $\{\mathcal{F}_{t_1},\mathcal{F}_{t_2}\}$ (blue line with circles) and $\{\mathcal{F}_{t_2},\mathcal{F}_{t_1}\}$ (red line with squares), where $\mathcal{F}_{t_1}$ and $\mathcal{F}_{t_2}$ are Hawkes processes with approximate power law kernel~\eqref{eq:pow_jpb_sum}.  The left panel corresponds to the case of concatenating processes with identical background activity $\mu_1=\mu_2=1$, but different branching ratios $n_1=0.5$ and $n_2$ spanning the interval $[0.05, 0.95]$. The right panel corresponds to the case of concatenating processes with identical branching ratio $n_1=n_2=0.5$, but different background activities $\mu_1=1$ and $\mu_2$ spanning the interval $[0, 2]$. All lines are obtained as averages over 100 different realizations. The confidence intervals are extremely narrow and are not presented: the maximum standard deviation of the estimation of $n$ in 100 different realizations is about $0.025$.} \label{fig:mixing}
\end{figure}

We illustrate this point with the following numerical simulation. We simulate two independent synthetic time series of the Hawkes process $\mathcal{F}_{t_1}$ and $\mathcal{F}_{t_2}$ with approximate power law kernel~\eqref{eq:pow_jpb_sum} for parameters $\mu_1=\mu_2=1$, $\theta_1=\theta_2=1$, $\tau_{0,1}=\tau_{0,2}=1$ second. Fixing the branching ratio $n_1=0.5$, we span the branching ratio $n_2$ within the interval $[0.05, 0.95]$. The length of realization is $T=10^5$ seconds, however as before we simulate realization on the interval $(0, 10^5+10^6]$ seconds and burn the initial period $(0, 10^6]$ to get rid of edge effects. Then, we concatenated the two time series to obtain continuous realizations $\{\mathcal{F}_{t_1},\mathcal{F}_{t_2}\}$ and $\{\mathcal{F}_{t_2},\mathcal{F}_{t_1}\}$ and calibrated Hawkes process with the same  approximate power law kernel~\eqref{eq:pow_jpb_sum} on the newly created realizations of length $T=2\cdot10^5$ seconds. Results are presented in fig.~\ref{fig:mixing} (left)\footnote{Note that, in contrast to figure~\ref{fig:robustness}, both lengths $T$ of the realizations and the initial burned period are larger, and the exponent $\epsilon$ is smaller, which results in more robust estimations.}. The branching ratio is always estimated to be larger than $0.5$, reaches the critical value $\hat n=1$ and even becomes super-critical ($\hat n>1$) when $n_2$ is significantly smaller than $n_1=0.5$. The dependence is especially steep for $n_2<0.5$, when concatenating realizations with $n_1=0.5$ and  $n_2=0.3$ results in 
the estimation $\hat n=0.8$. For $n_2=0.2$, the estimated branching ratio of the concatenated realization is critical: $\hat n=1$. 

Similarly, for two independent Hawkes process with identical branching ratio $n$, but with different background activity $\mu_1$ and $\mu_2$, concatenation results in a significant overestimation of the branching ratio $n$. Similarly to the previous simulation, we consider now independent synthetic time series of the Hawkes process $\mathcal{F}_{t_1}$ and $\mathcal{F}_{t_2}$ with identical branching ratio $n_1=n_2=0.5$, but with different values of the background activity: $\mu_1=0.5$ and $\mu_2$ spanning the interval $[0, 2]$. Results are presented in fig.~\ref{fig:mixing} (right). Similarly to the previous case, even small differences between $\mu_1$ and $\mu_2$ result in significant overestimations of $\hat n$. For example, a $40\%$ difference results in the estimation $\hat n=0.9$ when the true value is $n=0.5$; a $60\%$ difference results in estimating the critical value $\hat n=1$ for branching ratio.

The impact of the regime shift from one set of parameter values to another set of parameter values
is so strong that a strong upward bias of the branching ratio occurs even in total absence of 
clustering and of triggering. This is shown by using several independent Poisson processes with different intensities 
$\{\lambda_i\}$, and by generating a continuous time series of events with regime shifts from one intensity to the
next in subsequent time windows. By construction, such a time series should be characterized 
by a vanishing branching ratio as there is no triggering and no clustering.
Calibrating the Hawkes model on such a time series with any kind of kernel actually leads to quite
high values of $\hat n$, depending on the amplitude of 
different Poisson intensities $\{\lambda_i\}$. 
Keeping in mind that the variability of the real intensity over a two months period is extremely high 
as illustrated in Fig.~\ref{fig:regimes}, we expect a significant upward bias
in the estimation of $\hat n$ just as result of the spurious interpretation of clustering by the Hawkes
calibration exercise.

\begin{figure}[h!]
  \centering
  \includegraphics[width=0.9\textwidth]{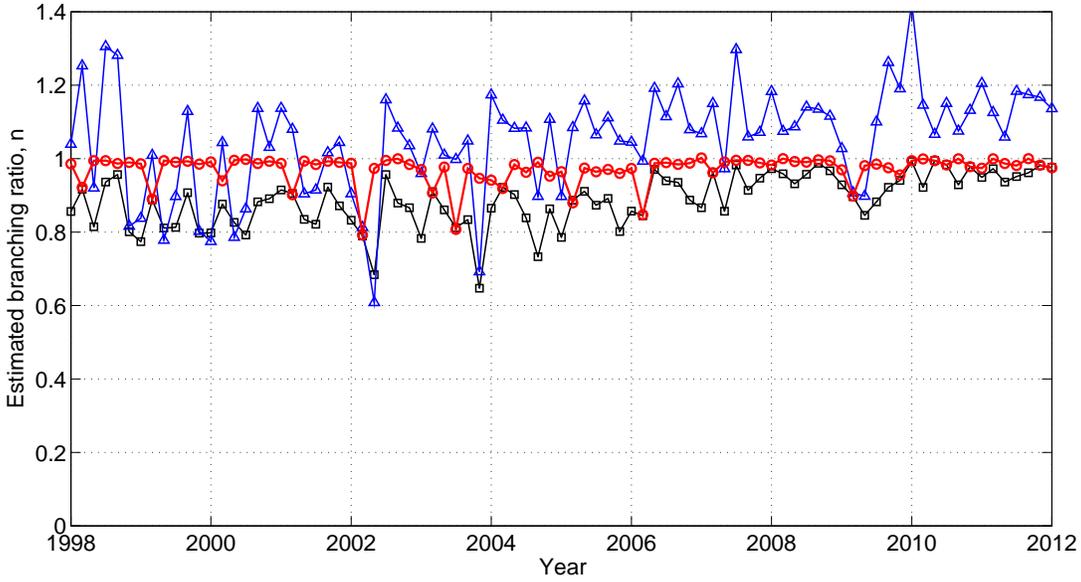}
  \caption{Results of the estimation of the branching ratio $n$ on the synthetic time-series concatenated from independent poisson processes with intensities taken from the real data (red curve). The black curve with squares corresponds to the case where days with trading durations smaller than 5 hours are ignored. The blue curve with triangles corresponds to the case where one continuous time series is constructed from independent Hawkes process with exponential kernel~\eqref{eq:exp} with $n=0.3$, $\tau=10$ seconds and $\mu$ proportional to the intensity taken from the real data.} \label{fig:poisson}
\end{figure}

In order to verify this prediction, we have constructed sets of synthetic time series in the manner of \citep{HardimanBouchaud2013} but, instead of real data, we have considered independent 
Poisson processes for each trading day, where the intensity $\lambda_i$ was estimated 
from the real data as equal to $\hat \lambda_i=N_i/T$ with $N_i$ being the number of observed 
mid-quote price changes in the $i$-th day, and $T=6.25$ hours is the duration of the Regular Trading Hours (RTH).
Fig.~\ref{fig:poisson} shows the estimated branching ratio in successive two-month windows from 1998 to 2012.
While the true branching ratio is zero by construction, the calibration by the Hawkes model gives
a value $\hat n$ hovering around the critical value $1$. Strikingly, while the truth is ``zero self-excitation'',
the calibration diagnoses `criticaility' over almost the whole 14 years interval. 
To test the robustness of this result, we also calibrated the Hawkes model on the time series
in which days where trading was stopped before 16:15 CDT were ignored
(some of these days correspond to the rollover dates). At these dates,
the total activity is extremely low (see for instance the first day of the third weeks in~Fig \ref{fig:regimes}) 
and such strong regime shift will drive the estimated $\hat n$ up. However, even by 
removing these days, the estimated branching ratio remains very high, above $0.8$ and close to the
critical value $1$ after 2007, as seen from the black continuous curve in fig.~\ref{fig:poisson}.

This synthetic non-stationary Poisson time series is too simple to 
embody the complex phenomenology reported in \citep{HardimanBouchaud2013},
such as the values of the exponent $\epsilon$ of the memory kernel which is estimated in the range $0<\hat\epsilon<1$
as compared to the structureless Poisson process. We have thus added some structure
in our synthetics by considering time series generated by the Hawkes model 
with an exponential kernel~\eqref{eq:exp}, $n=0.3$ (mild self-excitation), $\tau=10$ seconds 
and $\mu$ proportional to the intensity taken from the real data 
(low-activity days were ignored). The blue curve in Fig.~\ref{fig:poisson}
shows that the estimated branching ratio often becomes super-critical (${\hat n} >1$), i.e., very far from its
true value $n=0.3$. In this case, the exponent $\epsilon$ is estimated to be $0.1<\hat\epsilon<0.3$, 
which is similar to what is reported by \cite{HardimanBouchaud2013}.
The existence of regime shift thus strongly bias all parameters and leads to utterly
spurious conclusions.
 
Moreover, the residual analysis of these calibrations
cannot reject the null hypothesis of Poisson residuals~\eqref{eq:xi}. Hence, residual analysis
does not reject the null that time series truly generated by regime-switching non-homogeneous Poisson processes
would be generated by the Hawkes process at criticality, notwithstanding the 
fundamental differences between generating and calibrating models. The same lack
of rejection of the null is obtained when the true generating process is the regime-switching
Hawkes model with an approximate power law kernel~\eqref{eq:pow_jpb_sum}, $n=0.5$, $\tau_0=1$ second and $\epsilon=0.5$, which is undistinguishable according to the residual analysis from 
a stationary Hawkes process at or above criticality with branching ratio $\hat n\gtrsim1$ and
with the exponent $0.1<\hat\epsilon<0.3$.

Finally, we stress that the spurious criticality reported here is not unique to the approximate power law kernel~\eqref{eq:pow_jpb_sum}. In the presence of non-stationarity, a calibration using the Hawkes model with any kernel (including short-memory exponential kernel~\eqref{eq:exp}) will result in significant biases on the estimated branching ratio $n$.

\section{Conclusion}

We have presented a careful analysis of a set of effects that lead
to significant biases in the estimation of the branching ratio $n$, arguably the key parameter of the 
Hawkes self-excited Poisson process. The motivation of our study stems from the
meaning of $n$ as a direct measure of endogeneity (or reflexivity), since $n$
is exactly equal to the fraction of  the average number of endogenously generated events among all events 
\citep{Sornette2003geo,FilimonovSornette2012_Reflexivity} for stationary time series.
Concretely, the measure $n=0.7-0.8$ reported in our recent studies
\citep{FilimonovSornette2012_Reflexivity,FilimonovSornetteUNCTAD2012_Commodities}
means that 70 to 80\% of all trades in the  E-mini S\&P 500 Futures Contracts 
and in major commodity future contracts are due in recent years to past trades rather than to
external effects or exogenous news. This result has important implications concerning
the efficient market hypothesis and the stability of financial markets in the presence 
of increasing trading frequency and volume. The detailed and careful study of 
the possible biases attached to the estimation of $n$ has been additionally catalyzed by
the recent claim based also on the calibration of the Hawkes process that financial market have been continuously
functioning at or close to criticality ($n \simeq 1$) over the last decades \citep{HardimanBouchaud2013},
a result in contradiction with our other studies 
\citep{FilimonovSornette2012_Reflexivity,FilimonovSornetteUNCTAD2012_Commodities}.

Our overall conclusion is that calibrating the Hawkes process is akin to an excursion
within a minefield that requires
expert and careful testing before any conclusive step can be taken. We have identified
five main sources of generally upward biases for the branching ratio, which are arguably
present in high-frequency financial data. The size of the biases are found to be largely sufficient
to explain the discrepancy between our previous results and the claim of 
\cite{HardimanBouchaud2013}. We find in fact that all the biases are likely to be 
present in the analysis of \citep{HardimanBouchaud2013}, which can be seen as 
a showcase of the difficulties inherent in the calibration of the Hawkes process.

Our careful analysis of the microstructure of high-frequency financial data, the impact
of regular trading hours and overnight trading, the effect of latency of trade recording,
the grouping of trade timestamps, and their bundling suggest that most if not all 
studies published until now that have been concerned with inter-event times
should be revisited. In particular, our investigations suggest that 
the observations and estimations of heavy tailed distributions presented in other papers, 
the reports of long memory correlation functions and of multifractal scaling of the inter-event durations may be strongly biased by the bursty nature of the raw data
and may be subjected to the ``bundling'' effects and non-stationarity due to day-to-day regime shifts.

Given the many distortions occurring in trade recording by various stock market exchanges, 
robust choices ought to be made before any analysis is performed. In particular, our results presented here suggests that it is very difficult if not impossible to estimate reliably long-memory effects using high-frequency
financial data. In previous works \citep{FilimonovSornette2012_Reflexivity,FilimonovSornetteUNCTAD2012_Commodities},
we have bypassed this issue of long memory to focus solely on the branching ratio
at time scales of no more than tens of minutes, for which one can show that most of the
problematic biasing effects can be tamed.

On the conceptual level, reported long-term memory and critical branching structures of the triggering process are heavily biased by and can even result entirely from the non-stationarity in the high-frequency financial time series.
Such findings can result from the misuse of the mono-scaling statistical tools (such as correlation function and Hawkes process) applied to intrinsically multi-scaling generated data. With the increase of the 
duration of the time intervals beyond the intraday time scales for which approximate stationarity holds
to days and months, these analyses suffer from the problem of mixing different feedback mechanisms. The
existence of many different kind of strategies, from those used by the long-term investors to 
the behaviors of hedgers, chartists, high-frequency traders and other market participants, 
creates non-trivial multifractal scaling of the price time series~(see for instance~\cite{ArneodoMuzySornette1998}). Thus, the application of the Hawkes model with a single memory kernel on datasets at large time scales (weekly,  monthly, and so on) is not consistent with the complexity of the system. Here, one should consider some multifractal extension of the Hawkes model, for example in the spirit of Self-Excited Multifractal Process~\citep{FilimonovSornette2011_SEMF} that combines explicitly self-exciting feedback together with a nonlinear multi-scaling response function. However, no multifractal point process models have yet been introduced.

\section*{Acknowledgements}

We are grateful to Spencer Wheatley for helpful discussions and suggestions while preparing this manuscript as well as for pointing out the impact of outliers (section~\ref{sec:robustness}). We also thank Igor Artyukhin for the help in numerical tests. We are grateful to Stephen Hardiman and Jean-Philippe Bouchaud for many long and fruitful discussions while preparing the manuscript, as well as for pointing out some
inconsistencies and mistakes in preliminary versions of this article.

We would like to express our deep gratitude to professor Alexander Saichev,  for many fruitful discussions and collaborations on the critical regime of  branching processes. 
Professor Saichev, who passed away on 8 June 2013, has been an outstanding contributor to the general theory and to the practical applications of self-excited processes that we have been developing at ETH Zurich.

\clearpage

\end{document}